\documentclass[submission,PhysCodeb]{SciPost}
\usepackage{lineno}
\usepackage{graphicx}
\usepackage{subcaption}
\usepackage[percent]{overpic}
\usepackage{listings}
\usepackage{tabularx}
\usepackage{ltablex}
\keepXColumns
\usepackage{booktabs}
\usepackage{soul}
\usepackage{xcolor}
\usepackage{xltabular}
\usepackage{amsmath}
\usepackage{ulem}
\usepackage{xcolor}
\usepackage{listings}
\usepackage{hyperref}
\usepackage[section]{placeins}

\definecolor{JuliaPurple}{HTML}{AF00DB}
\definecolor{JuliaGreen}{HTML}{008000}
\definecolor{JuliaRed}{HTML}{A31515}
\definecolor{JuliaBlue}{HTML}{0000FF}
\definecolor{JuliaGray}{HTML}{808080}

\lstdefinelanguage{julia}{
    morekeywords={
        begin,while,for,in,return,break,continue,macro,quote,let,
        if,elseif,else,try,catch,finally,end,do,using,module,
        import,export,baremodule,struct,mutable,abstract,primitive,
        type,where,local,global,const,function,macro
    },
    morekeywords=[2]{
        Bool,Int,Int8,Int16,Int32,Int64,Int128,
        UInt,UInt8,UInt16,UInt32,UInt64,UInt128,
        Float16,Float32,Float64,BigInt,BigFloat,
        Complex,ComplexF16,ComplexF32,ComplexF64,
        Any,Nothing,Missing,String,Char,Symbol,
        Tuple,NamedTuple,Array,Vector,Matrix,Dict,Set,
        Pair,Task,IO,Regex,DataType,Type
    },
    sensitive=true,
    morecomment=[l]\#,
    morestring=[b]",
    morestring=[b]',
}

\hypersetup{
    colorlinks,
    linkcolor={red!50!black},
    citecolor={blue!50!black},
    urlcolor={blue!80!black}
}

\usepackage[bitstream-charter]{mathdesign}
\DeclareSymbolFont{usualmathcal}{OMS}{cmsy}{m}{n}
\DeclareSymbolFontAlphabet{\mathcal}{usualmathcal}

\fancypagestyle{SPstyle}{
    \fancyhf{}
    \lhead{\colorbox{scipostblue}{\bf\color{white}~SciPost Physics Codebases }}
    \rhead{{\bf\color{scipostdeepblue}~Submission }}
    
    \fancyfoot[C]{\textbf{\thepage}}
}

\begin{document}

\pagestyle{SPstyle}

\begin{center}
{\Large\textbf{\color{scipostdeepblue}{
QuantumBilliards.jl: An extensible, user friendly, high-performance Julia library for the large scale computation of spectra and eigenstates of quantum billiards and related problems
}}}
\end{center}

\begin{center}
\textbf{
Matic Orel\textsuperscript{1$\star$} and
Črt Lozej\textsuperscript{$\dagger$}
}
\end{center}

\begin{center}
{\bf 1} CAMTP -- Center for Applied Mathematics and Theoretical Physics,
University of Maribor, Maribor, Slovenia
\\[\baselineskip]
$\star$ \href{mailto:tensor1995@gmail.com}{\small tensor1995@gmail.com}\,,\quad
$\dagger$ \href{mailto:clozej@gmail.com}{\small clozej@gmail.com}
\end{center}

\begin{abstract}
We present \texttt{QuantumBilliards.jl}, an open-source Julia library providing
a suite of numerical solvers for the two-dimensional interior Helmholtz
eigenproblem with Dirichlet boundary conditions, a paradigmatic model in
quantum chaos. The library implements conventional wavenumber-sweep techniques
together with several accelerated eigensolvers: the Vergini--Saraceno scaling
method, the expanded boundary integral method (EBIM), Beyn's contour-integral
method, and the Compact Rational Krylov (CORK) method. It provides a common
framework for constructing billiard geometries, exploiting discrete
symmetries, computing eigenvalues and eigenstates, and treating smooth,
cornered, and multiply connected domains. The library further includes
reconstruction of eigenfunctions from boundary normal derivatives and the
computation of Poincar\'e--Husimi (PH) phase-space representations.

Available at: \hyperlink{lib}{https://github.com/Quantum-Chaos-Julia/QuantumBilliards.jl}.
\end{abstract}

\noindent\textcolor{white!90!black}{%
\fbox{\parbox{0.975\linewidth}{%
\textcolor{white!40!black}{\begin{tabular}{lr}%
  \begin{minipage}{0.6\textwidth}%
    {\small Copyright attribution to authors. \newline
    This work is a submission to SciPost Physics Codebases. \newline
    License information to appear upon publication. \newline
    Publication information to appear upon publication.}
  \end{minipage} & \begin{minipage}{0.4\textwidth}
    {\small Received Date \newline Accepted Date \newline Published Date}%
  \end{minipage}
\end{tabular}}
}}
}

\vspace{10pt}
\noindent\rule{\textwidth}{1pt}
\tableofcontents\thispagestyle{fancy}
\noindent\rule{\textwidth}{1pt}
\vspace{10pt}

\section{Introduction}
\label{sec:intro}

Mathematical billiards are some of the most well studied classical and quantum dynamical systems \cite{tabachnikovGeometryBilliards2005}. In spite of their apparent simplicity they reveal a surprisingly rich dynamical landscape. In the standard example we observe a (classical) point particle on a two-dimensional billiard table $\mathcal{B} \subset \mathbb{R}^2$. The particle may move freely inside the billiard table and is specularly reflected upon hitting the wall. Thus the dynamics can be described by two simple rules: the particle moves in straight lines and bounces off the wall with the angle of reflection equal to the angle of incidence. We may formalize the dynamics as a Hamiltonian system with
\begin{equation}
    H=\frac{\boldsymbol{p}^{2}}{2m}+V\left(\boldsymbol{r}\right)
    \end{equation}\label{eq:Hamiltonian}
where $p$ is the momentum, $m$ is the mass and the potential is
    \begin{equation}
    V\left(\boldsymbol{r}\right)=\begin{cases}
    0 & \boldsymbol{r}\in\mathcal{B},\\
    \infty & \boldsymbol{r}\notin\mathcal{B}.
    \end{cases}\label{eq:Potential}
    \end{equation}
Because of the simple dynamical rules one may reconstruct the complete trajectory of the particle from the wall collision sites and angles. The speed of the particle does not change the dynamics qualitatively as it only rescales the time. The standard approach is to study the dynamics as a discrete symplectic map in the Poincaré-Birkhoff coordinates $(q, p)=(s,\sin(\alpha))$, where $s$ is the arc-length measured counter-clockwise from some reference point on the billiard wall and $\alpha$ is the reflection angle. The phase space is a cylinder $\mathcal{M}=\left[0,\,\mathcal{L}\right]\times(-1,\,1)$. The fact that complete information of the dynamics is contained on the boundary will serve as inspiration for the numerical methods introduced in solving the quantum problem. The variable we control is the shape of the billiard table. The great wealth of dynamical phenomena we may observe with billiards is thus limited only by finding imaginative constructions of billiard tables. One may find tables with completely regular dynamics (circles, ellipses, rectangles...), chaotic billiards (like for instance the famous dispersive billiards of Sinai \cite{Sinai1970} and stadia of Bunimovich \cite{bunimovich1974ergodic}) as well as many examples of billiards with divided phase spaces (the mushroom billiards with only one island of stability \cite{Bunimovich_stadium,Bunimovich_ext_int_stickiness}, the Robnik billiards that interpolate between the circle and the limacon \cite{Robnik_Limacon}, lemon billiards \cite{HellerTomsovic_Lemon} with interesting special examples with small numbers of regular islands). Particularly interesting examples may be found among polygonal billiards that stradle the divide between chaos and integrability and explore intermediate levels of the ergodic hierarchy (for instance the mixing triangles \cite{CasatiProsen_Mixing_triangles} and the pseudointegrable Veech polygons \cite{Veech_Triangles,Pseudointegrable_triangles}). Other interesting dynamical regimes include chaotic billiards with slow decay of correlations and the stickiness phenomenon (a very typical intermittent type of motion where chaotic trajectories spend a lot of time in certain regions). Furthermore many generalizations have been considered like magnetic billiards \cite{Hornberger_Smilansky}, dissipative and open billiards \cite{openbilliards} time-dependent billiards to name a few \cite{timedepbilliards}).

The object of this paper and indeed the QuantumBilliards.jl numerical library is the quantum generalization of mathematical billiards. In this setting we quantize the Hamiltonian \eqref{eq:Hamiltonian} by the standard procedure of replacing the momentum with the quantum operator 
\begin{equation}
    H=-\nabla^{2}+V\left(\boldsymbol{r}\right),
    \end{equation}\label{eq:HamiltonianQuantum}
where we customarily set the units so that $\hbar^{2}/2m=1$ and the potential is the same as above \eqref{eq:Potential}. The objective thus becomes solving the Schrödinger equation, by solving the stationary Helmholtz problem and finding the spectrum (energy levels or wavenumbers $E=k^{2}$) and eigenfunctions inside the bounded domain $\mathcal{B}$ with Dirichlet boundary conditions. Quantum billiards are a cornerstone in the study of quantum
chaos~\cite{Stoe,Haake,Rob2020,Rob2023}. Because the classical dynamics does not depend on the energy, the semiclassical limit in quantum billiards is especially clean, namely $k \to \infty$ (equivalently $E \to \infty$ at fixed $\hbar$, or $\hbar \to 0$ at fixed $E$), meaning $k L \gg 1$ for a characteristic length $L$ which makes them an ideal testing ground for ideas and numerical verification of theoretical conjectures. Many of the most prominent results in quantum chaos have been shown or indeed discovered in quantum billiards like the universal spectral statistics of chaotic ~\cite{Cas1980,BGS1984,Bro1973,Izr1989} and integrable systems \cite{BerryTabor,Veble_Robnik_integrable}, quantum ergodicity \cite{Shnirelman}, intermediate \cite{BogomolnyGerlandSchmit1999_Spectral_Statistics,lozej2022triangles, lozej_intermediate_2024}, and mixed-type \cite{BerryRobnik1984_their_spectral_statistics,BarBet2007} spectral statistics,  quantum scars \cite{Heller1984} and superscarrs \cite{bogomolnyStructureWaveFunctions2004},  eigenstate
localization~\cite{BLR2020,batistic_robnik_localization_eigenstates, lozejPhenomenologyQuantumEigenstates2022, lozejSpectralFormFactor2023}, quantum sticky-states \cite{KetzmerickEtAl2000_Sticky_quantum_states, LozejThesis, crt_stickiness}, and
semiclassical correspondence~\cite{percival1973,Rob1998,Rob2023A}. 

\begin{figure}[htbp]
    \centering
    \includegraphics[width=1.0\linewidth]{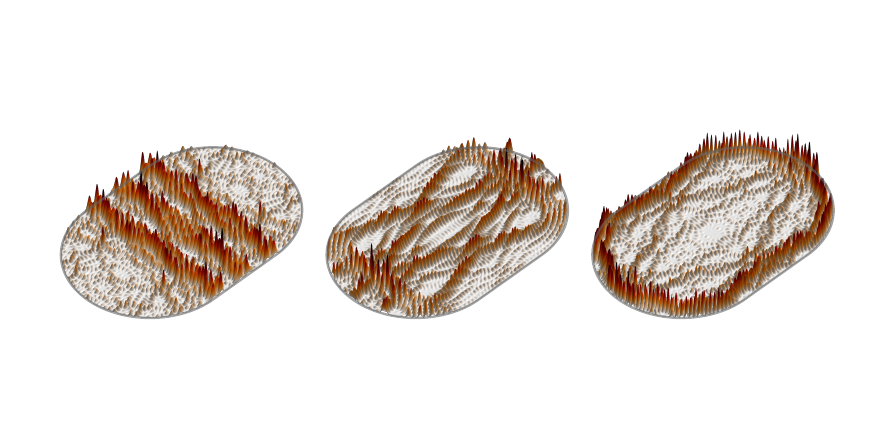}
    \caption{Examples of scarred eigenstates of the stadium billiard with half width equal to 1. The wavenumbers are $k=74.1838018$, $74.22509066$, $74.26339429$.} 
\end{figure}

Quantum billiards have been realized on a wide range of experimental platforms.
Because the Helmholtz and Schr\"odinger equations coincide in two dimensions,
flat metallic microwave cavities, including superconducting ones, serve as
analogue billiards whose spectra and field patterns can be measured directly
\cite{Stockmann1990,Sridhar1991,Graf1992,Alt1995,So1995,Stoffregen1995}.
Elastic and acoustic resonators such as machined metal blocks offer a similar
wave-based route \cite{Weaver1989,Ellegaard1995}. Electronic billiards are
realized by confining surface-state electrons with atomic corrals
\cite{Crommie1993} or by defining ballistic cavities in semiconductor
heterostructures \cite{Jalabert1990,Marcus1992,Chang1994,Ketzmerick1996,Sachrajda1998},
where Coulomb blockade offers a further probe \cite{Jalabert1992,Folk1996,ChangA1996,Alhassid2000}.
In optics, shaped dielectric microcavities and microlasers act as open billiards
\cite{Nockel1997,Gmachl1998}, and light-sheet walls have been used to build
billiards for cold atoms \cite{Friedman2001,Milner2001,Stoe,Richter1998,experiment_microwave_tunneling,Jiang_laser,Harayama_17,
stockmann_microwave_billiards_2022}.

Many contemporary studies require considerably more than isolated low-lying
eigenvalues. Statistical and semiclassical investigations may involve large
ensembles of highly excited eigenstates together with accurate boundary
functions, given for Dirichlet billiards by the normal derivative
of the wavefunctions, and phase-space representations such as
Poincar\'e--Husimi functions~\cite{Hus1940,TV1995,Backer2003}. The boundary
function is particularly important in boundary-integral formulations, where
it provides a natural representation of a Dirichlet eigenstate
~\cite{backer_BIM,kress_book}. These requirements become increasingly
demanding in the semiclassical regime: the number of required collocation points on the boundary needed to resolve it grows with wavenumber, dense discretizations
become computationally expensive~\cite{lozej2022triangles}, and geometric
singularities such as corners require specialized numerical treatment
~\cite{kress,rcip_helsing}.

The numerical solution of the Helmholtz problem has a long history ranging back to the 1960s \cite{FoxHenriciMoler1967_Particular_Solutions_Method,betcke} (as a curiosity the MATLAB logo is a plot of the first eigenfunction of the L-shaped membrane equivalent to a polygonal quantum billiard) and 
a variety of numerical approaches have been developed for the Helmholtz
eigenvalue problem. Basis-expansion and method-of-particular-solutions
approaches represent the wavefunction using functions which already satisfy
the Helmholtz equation in the interior~\cite{Heller1984,betcke} and minimize tension norms on the boundary to fulfill the boundary conditions. However these methods are computationally very expensive as they require multiple matrix diagonalizations per found eigenvalue and states may also be missed if local minima are not resolved. The
Vergini--Saraceno scaling method provides an efficient extension of the basis expansion and minimization idea. By expanding the tension norms around a reference wavenumber many nearby eigenvalues and eigenstates may be extracted from a
single calculation without missing levels~\cite{VerSar1995,Barnett2000,BarBet2007}. 
Boundary-integral equation (BIE) methods instead reduce the problem to the
boundary of the billiard. After discretization this leads to a nonlinear matrix
eigenvalue problem~\cite{backer_BIM,kress_book}. Local expansion of the
boundary operator as a generalization of the scaling method idea results in the expanded boundary integral method
~\cite{VebProRob2007,dietz_neutrino_expanded}, while contour-integral methods
such as Beyn's method provide a global approach for extracting multiple
eigenvalues enclosed by a contour in the complex wavenumber plane
~\cite{beyn_contour,sakurai_contour}.

Several reusable software packages provide numerical machinery relevant to
these problems, although their scope and intended applications differ
considerably. MPSpack~\cite{barnett_mpspack} is a MATLAB toolbox specifically
developed for two-dimensional Laplace and Helmholtz problems and provides
both method-of-particular-solutions and some boundary-integral approaches. General-purpose integral-equation
frameworks such as \texttt{Inti.jl}~\cite{Inti} and
\texttt{chunkIE}~\cite{ChunkIE} provide
numerical tools for boundary-integral operators over a broader
class of boundary-value problems. In three dimensions,
\texttt{Bempp}~\cite{bempp} provides a general boundary-element framework
including Helmholtz layer-potential operators. A related Julia package, 
\texttt{WavesAndEigenvalues.jl} \cite{WavesAndEigenvalues}, provides generic
solvers for nonlinear eigenvalue problems. These packages therefore provide
overlapping components of the required numerical machinery, but target different 
geometries, discretizations, physical problems, and stages of the computational workflow.
Large-scale quantum-billiard calculations typically require several of these
capabilities simultaneously. The availability of several independent
eigensolvers within the same framework is also useful: different numerical 
formulations are advantageous in different parameter regimes, while agreement 
between independent methods provides an important means of validating computed spectra.

In this work we present \texttt{QuantumBilliards.jl}, an open-source Julia
framework designed specifically to solve the whole  two-dimensional quantum-billiard
workflow  for general geometries. It combines basis-expansion methods~\cite{Barnett2000,betcke},
boundary-integral formulations~\cite{backer_BIM,kress_book}, and several
accelerated eigensolvers within a common interface. These include the
Vergini--Saraceno scaling method, the expanded boundary integral method
~\cite{VebProRob2007,dietz_neutrino_expanded}, Beyn's contour-integral method
~\cite{beyn_contour}, and a Compact Rational Krylov \cite{CORK} (CORK) method for the
nonlinear boundary-integral eigenvalue problem. The same framework supports
smooth, cornered, symmetry-reduced, and multiply connected geometries and
connects the resulting spectra directly to eigenfunction and
phase-space analysis. We may mention that a precursor of this package, 
implementing the basis solvers in Python and C++ are available at \cite{BatisticQuantumBilliards,LozejQuantumBilliards,BarnettSoftware}.

The emphasis of the library is therefore not on a single numerical
formulation, but on providing a common computational environment in which
different high-frequency quantization methods can be applied to the same
billiard geometry and their results independently cross-validated.
The Julia programming language was chosen for its best in class combination of speed and numerical efficiency and user friendliness. The multiple dispatch system allowed for a user friendly API (application programming interface) design without sacrificing any numerical performance. The package is designed to serve as a ground layer for further extensions to related problems that can be represented with different boundary conditions such as for instance dielectric micro-cavities as well as billiard state time evolutions planned in the future. We hope this package will extend the already impressive collection of scientific computing libraries such as DynamicalSystems.jl \cite{Datseris_Dynamical_Systems} and DifferentialEquation.jl \cite{Rackauckas_Differential_Equations} which served as inspiration.

\section{Solving the Helmholtz problem}
\label{sec:eigensolvers}

The main numerical problem considered throughout \texttt{QuantumBilliards.jl} is the
interior Dirichlet Helmholtz eigenproblem in a 2-dimensional domain $\mathcal{B}$
\begin{equation}
    (\nabla^2+k^2)\psi(\mathbf r)=0,
    \qquad \mathbf r\in\mathcal{B},
    \label{eq:helmholtz}
\end{equation}
with
\begin{equation}
    \psi(\mathbf r)=0,
    \qquad \mathbf r\in\partial\mathcal{B}.
    \label{eq:dirichlet}
\end{equation}
The objective is to determine the discrete eigenwavenumbers $k_n$ for which
a non-trivial solution $\psi_n$ exists. We denote the eigenenergy $E_n=k_n^2$. The library approaches this problem
through two complementary numerical representations. In the \textit{basis methods} the
wavefunction is expanded in a (typically overcomplete) set of functions which already satisfy
Eq.~\eqref{eq:helmholtz}, leaving the boundary condition to be imposed
numerically. Different basis sets may also be specially tailored to satisfy certain geometries or boundary conditions (on part of the billiard wall) automatically. In \textit{boundary-integral methods} (BIM)the two-dimensional problem is
instead reduced to a nonlinear operator equation on the one-dimensional
boundary. In both cases the eigenstates may then be found by minimizing a generalized non-negative tension $\tau(k)$, which should approach zero at eigenwavenumbers. 
Different minimization norms or tensions may be used to produce different variants of the BIM and basis methods, which may also be seen as dual formulations (for a detailed analysis see Lepore's PhD thesis \cite{Lepore_thesis}). However the main weakness of applying these methods directly is the need for sweeping the desired wavenumber intervals and finding the local minima which may also easily be missed. This is typically very numerically expensive as it requires repeatedly constructing and solving a dense
matrix problem at individual trial wavenumbers and quickly becomes untenable. We will collectively refer to these methods as \textit{sweep solvers}. However, valuable information may still be gained about the performance of the different basis constructions and convergence around the minima and this is why we chose to include them in the library. 

The main workhorse of the spectral computations will be the \textit{accelerated solvers}. The original idea came from Vergini and Saraceno \cite{VerSar1995} as an accelerated version of Heller's plane wave decomposition method \cite{Heller1984} (this is a variant of the basis method where we use real plane waves as the set of basis functions). By exploiting a scaling relation in the basis functions and using a Taylor expansion of a specially weighted tension norm it is possible to find multiple wavenumbers within a small interval around a reference $k$ with a single step. This massively improves the numerical efficiency and has proven to be one of the most practical methods for large scale numerical studies of quantum billiard spectra and eigenstates \cite{LozejCasatiProsen2022_Triangle_large_spectrum,OreLozRobYan2025}. The same basic idea of Taylor expansion around a reference wavenumber was later extended to BIM by Veble et. al. \cite{VebProRob2007}. We will collectively refer to this kinds of methods as \textit{accelerated solvers}. 

Recently, Beyn's method using this library has been applied to quantum billiards \cite{OrelRobnik2026C3,Orel2026Hyperbolic}. An implementation using higher order moments of the inverse of the Fredholm matrix was used in \cite{YiDietzHanRyu2025_Sakurai_high_moments}. This library contains the streamlined version of the implementation of Beyn's method from the aforementioned study as well as a new approach following the CORK algorithm that may provide further performance gains. In the following sections we explore each solver that is implemented in our library and compare their strengths and weaknesses. 

Future development will extend both the numerical methods and the physical problems supported by the framework. Unexplored future directions include high-frequency spectral-flow formulations based on Neumann-to-Dirichlet maps~\cite{barnett_NtD}, extensions to problems governed by different Green functions, such as billiards on the hyperbolic metric
~\cite{hyperbolic_triangle_schmit,
hyperbolic_billiards_book,Lepore_Heller_hyperbolic_triangle}
and billiards in a homogenous magnetic field
~\cite{Hornberger_Smilansky} (see Appendix.\ref{sec:appendix_future}) and relativistic neutrino
billiards~\cite{dietz_neutrino_expanded}. On the numerical side, further developments include recursively compressed inverse preconditioning for the treatment of corners ~\cite{rcip_helsing} and charge bases adapted to corner singularities \cite{Bremer_Rokhlin_corner_charges}. An interesting but out-of-scope approaches are QBX and QB2X~\cite{QBX,QB2X} for accurate evaluation of layer potentials on and near the boundary.

\subsection{Helmholtz basis formulation}
\label{sec:basis_formulation}

In the basis formulation the wavefunction is approximated by

\begin{equation}
    \psi(\mathbf r;k)
    =
    \sum_{j=1}^{M}c_j\phi_j(\mathbf r;k),
    \label{eq:basis_expansion}
\end{equation}

\noindent where every basis function satisfies

\begin{equation}
    (\nabla^2+k^2)\phi_j(\mathbf r;k)=0,
    \label{eq:basis_helmholtz}
\end{equation}
in free space.
\noindent The Helmholtz equation is therefore satisfied by construction and the
numerical problem reduces to finding a non-trivial coefficient vector
$\mathbf c$ for which

\begin{equation}
    \sum_{j=1}^{M}c_j\phi_j(\mathbf r;k)=0,
    \qquad
    \mathbf r\in\partial\mathcal{B}.
    \label{eq:basis_boundary_problem}
\end{equation}

\begin{figure}[htbp]
    \centering
    \includegraphics[width=0.8\linewidth]{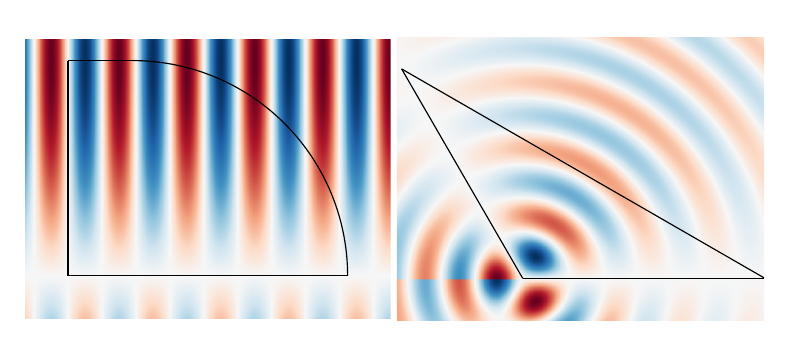}
    \caption{Examples of Helmholtz basis functions implemented in
    \texttt{QuantumBilliards.jl}. Left: a real plane-wave basis function for the
    stadium billiard. Right: a corner-adapted Fourier--Bessel basis function for
    a triangular billiard, centered at a corner so that the boundary
    conditions are incorporated directly into the basis. Eigenfunctions
    are constructed as linear combinations of such functions according to
    Eq.~\eqref{eq:basis_expansion}. For high-accuracy calculations, additional evanescent plane waves may be required near parts of the boundary that are difficult to resolve.}
    \label{fig:bases}
\end{figure}

\noindent There are different possibilities for choosing the basis functions. It is often advisable (or in most case necessary for convergence) to select basis functions that automatically fulfill the boundary conditions on part of the billiard wall. Historically, real plane waves were used initially, since elementary functions are easy to compute and thus this was known as the \textit{plane-wave decomposition method} \cite{Heller1984} (PWDM). However, due to the finite-precision limitations in actual computations the convergence of this basis is in practice limited to mostly smooth and convex geometries although convergence is not guaranteed even in this case. We implement the \textit{real-plane wave basis} as
separable Helmholtz solutions

\begin{equation}
    \phi_j(x,y;k)
    =
    F_x\!\left(kx\cos\theta_j\right)
    F_y\!\left(ky\sin\theta_j\right),
    \label{eq:real_plane_waves}
\end{equation}

\noindent where $F_x,F_y\in\{\cos,\sin\}$. For reflection-symmetric billiards the
choice of sine or cosine incorporates the selected parity directly into the
basis. The main advantage of this choice is the speed of basis evaluations. The \textit{Fourier-Bessel basis}~\cite{Barnett2000,BarBet2007} is another popular choice. In local polar coordinates $(r, \theta)$ centered at some origin, 
\begin{equation}
    \phi_j(r,\theta) = J_j(kr)\,\begin{cases}\cos(j\theta)\\ \sin(j\theta)\end{cases},     \qquad
    j=0,1,2,\ldots
\end{equation}
For geometries containing a distinguished corner of interior angle $\alpha$, like triangles or mushrooms,
the \textit{corner-adapted} variant is especially useful

\begin{equation}
    \phi_j(r,\varphi;k)
    =
    J_{\nu j}(kr)\sin(\nu j\varphi),
    \qquad
    \nu=\frac{\pi}{\alpha},
    \qquad
    j=1,2,\ldots,
    \label{eq:corner_fourier_bessel}
\end{equation}

\noindent which satisfies the Dirichlet condition on the two rays forming the corner.
Examples of the two basis families and representative eigenfunctions obtained
from them are shown in Fig.~\ref{fig:bases}.

To evaluate the boundary condition we introduce the following bilinear boundary form 
\begin{equation}
    \oint_{\partial\mathcal{B}}|\psi|^2\,ds = \mathbf c^\dagger F(k)\,\mathbf c
    \label{eq:tension_norm}
\end{equation}
Additionally we introduce the bilinear
normalization form 
\begin{equation}
\int_\mathcal{B} |\psi|^2\,d^2r = \mathbf c^\dagger N(k)\,\mathbf c.
\label{eq:psm_normalization}
\end{equation}

The eigenvalues are found where the boundary form is minimal subject to a fixed value of the normalization which gives a generalized eigenproblem
\begin{equation}
F(k)\,\mathbf c=\tau\,N(k)\,\mathbf c
\end{equation}
which in practice means solving a singular value matrix problem to find the vector that minimizes the tension at a given wavenumber
\begin{equation}
    \tau(k)
    =
    \min_{\mathbf c\neq0}
    \frac{\|F\mathbf c\|_2}
         {\|N\mathbf c\|_2}.
    \label{eq:dm_tension}
\end{equation}
The \textit{particular solutions method} ~\cite{betcke,BarBet2007} (PSM) solves this minimization problem directly.
\noindent Writing the boundary norm in the finite basis expansions we obtain the matrix elements
\begin{equation}
F_{ij}(k)=\oint_{\partial\mathcal{B}}\overline{\phi_i(\mathbf r;k)}\,\phi_j(\mathbf r;k)\,ds
\end{equation}
and after discretization on the boundary points $\mathbf r_l$, $l=1,\dots,M$, with quadrature weights $w_l$ \cite{Monte_Carlo}
\begin{equation}
{F_{ij}(k)\approx\sum_{l=1}^{M} w_l\,\overline{\phi_i(\mathbf r_l;k)}\,\phi_j(\mathbf r_l;k)}.
\label{eq:F_matrix}
\end{equation}
Since the boundary norm is evaluated only on the boundary points the number of collocation points needed for convergence will scale with the length of the boundary rather then the area. In case the billiard boundary is smooth, even a simple trapezoidal quadrature rule provides exponential convergence since the basis functions are analytic on the boundary. The normalization is evaluated using points sampled inside the domain (the specific quadrature role does not matter, as using a rank--revealing QR decomposition \cite{RRQR} on $N_{ij}(k)$ in \eqref{eq:psm_normalization} removes near null (correlated) directions). 

The idea behind the PWDM, or as we will refer to it more generally, the  \textit{decomposition method} (DM) is to substitute the second form with a boundary integral as well. Originally Heller proposed to evaluate the wavefunction at a single random point inside the billiard to fix the solution to finite value, but an unlucky coincidence where the point is near a nodal line where the wavefunction is close to 0 may produce failures. We use Barnett's formulation \cite{barnett_dissipation} where we translate the area integral to another boundary integral by using the Rellich identity \cite{rellich1940}. To derive it we first require the divergence theorem. 
For any smooth vector field $\mathbf V$,
\begin{equation}
\int_\mathcal{B}\nabla\cdot\mathbf V\,d^2r=\oint_{\partial\mathcal{B}}\mathbf V\cdot\hat{\mathbf n}\,ds.
\end{equation}
Using the fact that all solutions satisfy the Helmholtz equation \eqref{eq:helmholtz}, and standard vector identities we may show that 
\begin{equation}
\nabla\cdot\mathbf V=k^2|\psi|^2,
    \qquad
\mathbf V=\mathrm{Re}\big[(\mathbf r\cdot\overline{\nabla\psi})\,\nabla\psi\big]-\tfrac12\,\mathbf r\,|\nabla\psi|^2+\tfrac12k^2\,\mathbf r\,|\psi|^2 .
\end{equation}
Then we decompose the vector field into normal and tangential parts 
\begin{equation}
\begin{aligned}
\mathbf r\cdot\nabla\psi &= r_n\partial_n\psi+r_t\partial_t\psi,
\qquad
|\nabla\psi|^2 = |\partial_n\psi|^2+|\partial_t\psi|^2,\\
\mathbf V\cdot\hat{\mathbf n}
&= \tfrac12 r_n\big(|\partial_n\psi|^2-|\partial_t\psi|^2+k^2|\psi|^2\big)
+r_t\,\mathrm{Re}\big(\overline{\partial_t\psi}\,\partial_n\psi\big).
\end{aligned}
\end{equation}
and therefore

\begin{equation}
\int_\mathcal{B}|\psi|^2\,d^2r=\frac{1}{k^2}\oint_{\partial\mathcal{B}}\Big[\tfrac12 r_n\big(|\partial_n\psi|^2-|\partial_t\psi|^2+k^2|\psi|^2\big)+r_t\,\mathrm{Re}\big(\overline{\partial_t\psi}\partial_n\psi\big)\Big]ds .
\end{equation}
In case of billiard eigenstates, Dirichlet boundary conditions apply and we may further simplify the expression since $\psi=\partial_t\psi=0$ on $\partial\mathcal{B}$ and obtain the Rellich identity 
\begin{equation}
    \int_\mathcal{B}|\psi|^2\,d^2r=\frac{1}{2k^2}\oint r_n|\partial_n\psi|^2ds. \label{eq:rellich}
\end{equation}
The quantity $u(s)=\partial_n\psi(\mathbf r(s))$ is the outward normal derivative of $\psi$ is known as the boundary function~\cite{backer_boundary,Backer2003} and as we will see later it contains all the information that is needed to reconstruct the wavefunction in the billiard interior in complete analogy to the classical case. Applied to the basis, this gives the simplified normalization matrix
\begin{equation}
N_{ij}(k)=\frac{1}{2k^2}\oint_{\partial\mathcal{B}}r_n\,\overline{\partial_n\phi_i}\,\partial_n\phi_j\,ds\;\approx\;\frac{1}{2k^2}\sum_{l=1}^M w_l\,r_{n,l}\,\overline{\partial_n\phi_i}(\mathbf r_l)\,\partial_n\phi_j(\mathbf r_l),
\label{eq:N_matrix}
\end{equation}
where the derivatives of the basis functions may all be evaluated analytically and only once at each boundary point in the case of symmetric construction. Both formulations therefore reduce the eigenvalue search to locating minima of \eqref{eq:dm_tension} which is a scalar function of $k$.

The principal numerical parameters are the size of the basis and the number and distribution of collocation points on the boundary. 
The basis dimension is chosen approximately as
\begin{equation}
    M
    \simeq
    b_{\rm basis}\frac{kL}{2\pi},
    \label{eq:basis_size}
\end{equation}
where $L$ is the relevant boundary length and $b_{\rm basis}$ corresponds
to the basis dimension oversampling parameter. The boundary discretization is scaled
analogously %
\begin{equation}
    N
    \simeq
    b_{\partial\mathcal{B}}\frac{kL}{2\pi},
    \label{eq:bim_size}
\end{equation}
These two parameters are
shared by both the DM and Particular solutions method PSM formulations. The PSM additionally uses
a scaling factor to control the number of interior
points of the boundary. Increasing these resolution parameters and verifying
the stability of the resulting eigenvalues provides the basic convergence
test for basis calculations. Crucially, the implementation removes numerically near-dependent basis directions automatically; the corresponding rank reduction is discussed in Appendix~\ref{sec:vs_details}.

\subsection{Boundary-integral formulation}
\label{sec:bim_formulation}

Boundary-integral methods instead use the free space Green's function to reduce Eqs.~\eqref{eq:helmholtz} and
\eqref{eq:dirichlet} to an equation involving only data on the billiard boundary
$\partial\mathcal{B}$~\cite{backer_BIM,kress_book}. By definition 
the Green's function $G_k(\mathbf r,\mathbf r')$ solves the Helmholtz equation with a point source:
\begin{equation}
(\nabla^2+k^2)\,G_k(\mathbf x,\mathbf y)=-\delta(\mathbf x-\mathbf y)
\label{eq:green_def}
\end{equation}
The outgoing free-space Green's function in two dimensions is

\begin{equation}
    G_k(\mathbf x,\mathbf y)
    =
    \frac{i}{4}H_0^{(1)}(kr),
    \qquad
    r=|\mathbf x-\mathbf y|,
    \label{eq:helmholtz_green}
\end{equation}

\noindent and its outward normal derivative with respect to the source point is

\begin{equation}
    \partial_{n_{\mathbf y}}G_k(\mathbf x,\mathbf y)
    =
    \frac{ik}{4}H_1^{(1)}(kr)
    \frac{(\mathbf x-\mathbf y)\cdot\mathbf n_{\mathbf y}}{r}.
    \label{eq:helmholtz_green_normal}
\end{equation}
Green's second identity applied to a Helmholtz solution $\psi$ and $G_k$ gives for $\mathbf x$ inside the billiard

\begin{equation}
\psi(\mathbf x)=\oint_{\partial\mathcal B}\big[G_k(\mathbf x,\mathbf y)\,\partial_n\psi(\mathbf y)-\psi(\mathbf y)\,\partial_{n_y}G_k(\mathbf x,\mathbf y)\big]ds_{\mathbf y}.
\end{equation}

\noindent These kernels define the single- and double-layer operators

\begin{equation}
    (S(k)\mu)(\mathbf x)
    =
    \int_{\partial\mathcal{B}}
    G_k(\mathbf x,\mathbf y)\mu(\mathbf y)\,ds_{\mathbf y},
    \qquad
    (D(k)\mu)(\mathbf x)
    =
    \int_{\partial\mathcal{B}}
    \partial_{n_{\mathbf y}}G_k(\mathbf x,\mathbf y)
    \mu(\mathbf y)\,ds_{\mathbf y},
    \label{eq:layer_operators}
\end{equation}
where $\mu$ is the layer density. The two kernels both solve the Helmholtz equation off the boundary curve (inside and outside the billiard) but their behaviour differs between the two sides. Taking boundary limit approaching from inside or outside the billiard the single layer potential is continuous, whereas the double layer potential has a jump. This is known as the jump relations \cite{kress_book}.  The limit results in a nonlinear boundary eigenproblem for the Fredholm matrix 
of the form

\begin{equation}
    A(k)\mu=0.
    \label{eq:bim_nep}
\end{equation}

After Nystr\"om discretization,
$A(k)\in\mathbb C^{N\times N}$, an eigenvalue of \eqref{eq:helmholtz} is characterized by loss of invertibility of $A(k)$. A robust sweep tension scalar function is therefore the smallest singular value~\cite{backer_BIM} (more robust than calculating the determinant of $A(k)$ where its eigenvalues become very large and the minimum can span large ranges),

\begin{equation}
    \tau_{\rm BIM}(k)
    =
    \sigma_{\min}[A(k)].
    \label{eq:bim_tension}
\end{equation}

\noindent Two closely related boundary operators are provided. The
\textit{double-layer potential solver} (DLP) uses \cite{backer_BIM}

\begin{equation}
    A_{\rm DLP}(k)=I-2D(k),
    \label{eq:dlp_operator}
\end{equation}

\noindent whereas the \textit{combined field-integral equation solver} (CFIE) combines
single- and double-layer contributions~\cite{kress},
\begin{equation}
    A_{\rm CFIE}(k)
    =
    I-2D(k)-2i\eta S(k).
    \label{eq:cfie_operator}
\end{equation}
The factor of $2$ corresponds to a rescaling of the boundary integral
equation in which the $\tfrac{1}{2}I$ jump term becomes $I$. The coupling
parameter is chosen as $\eta=k$, following the near-optimal choice proposed
by Kress~\cite{Kress_optimal_coupling}. The combined-field formulation
avoids the spurious resonances associated with the pure double-layer
formulation~\cite{backer_BIM}. The DLP is therefore a convenient and
efficient formulation when its auxiliary resonances are not problematic,
whereas the CFIE provides a more robust formulation at the cost of additional
special-function evaluations.

For multiply connected billiards, we provide a composite solver that allows each
connected boundary component to use its CFIE discretization while
assembling all self- and mutual connected boundary interactions into one globally coupled operator. This is useful when different components (holes) require different quadrature or corner treatment.

The principal numerical parameter of both the DLP and CFIE formulations is
the boundary resolution \eqref{eq:bim_size}. Its appropriate value depends on the geometry
and quadrature and should be established by convergence. For piecewise-smooth
boundaries, the boundary grading provides an additional discretization
control near corners, as discussed in Appendix~\ref{sec:geometry}. Both
solvers can additionally exploit discrete symmetries by specifying
the symmetry of the wavefunction, thereby restricting the calculation
to a selected symmetry sector. The construction and interpretation of these
sectors are described in Appendix~\ref{sec:geometry}.

\begin{figure}[h!]
    \centering
    \includegraphics[width=\linewidth]{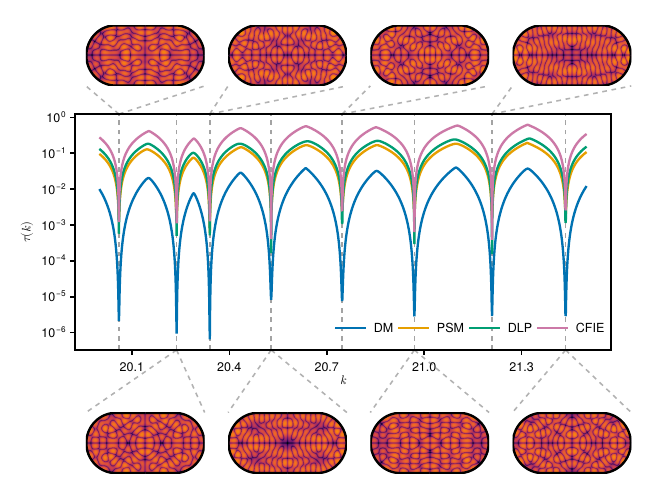}
    \caption{Direct spectral sweep for the stadium billiard in the
    $(-1,-1)$ reflection-symmetry sector over $k\in[20,25]$. The DM, PSM,
    DLP, and CFIE tensions are defined differently and therefore have
    different absolute scales, but their minima occur at the same physical eigenwavenumbers. The corresponding eigenfunctions are shown at various sweep tension minima and illustrate the different spatial structures of states. The wavefunction color-plots are shown in the logarithmic scale, with the black lines being the nodal lines.}
    \label{fig:sweep_all}
\end{figure}

Figure~\ref{fig:sweep_all} shows a comparison sweep over $k
$ for all the basis and BIM solver formulations in the stadium billiard in the odd-odd symmetry sector. Their absolute
magnitudes should not be compared because Eqs.
\eqref{eq:dm_tension}, and \eqref{eq:bim_tension} define different
scalar tension functions $\tau$. Their minima, however, identify the same
solution. The corresponding wavefunction is also shown in the logarithmic colorplots to illustrate the nodal structure of each state.

\subsection{Kress quadrature scheme}
\label{sec:kress_quadrature}

The single- and double-layer kernels in Eq.~\eqref{eq:layer_operators} have
different behaviour on the diagonal after discretization. The single-layer kernel has a
logarithmic singularity, whereas the double-layer kernel is continuous for a
smooth boundary, although its derivatives have logarithmic singularities \cite{kress}.
For numerical accuracy it is therefore advantageous to separate the
logarithmic contribution in both cases~\cite{kress,kress_book}.

Let $\mathbf{x}(t)$, $t\in[0,2\pi)$, be a smooth periodic
parametrization of the boundary, with $t$ and $\tau$ denoting the
parameters of the target and source points, respectively. The
kernels \eqref{eq:layer_operators} can then be written in the form

\begin{equation}
    K(t,\tau)
    =
    K_1(t,\tau)
    \log\left[
        4\sin^2\left(\frac{t-\tau}{2}\right)
    \right]
    +
    K_2(t,\tau),
    \label{eq:kress_split}
\end{equation}

\noindent where $K_1$ and $K_2$ are smooth periodic functions (or using $\log{|t-\tau|}$ for doing high order Gauss-Legendre logarithmic product quadrature \cite{SmithSloan1980_Gauss_Legendre}). Kress product
quadrature scheme~\cite{kress,kress_book} treats the logarithmic contribution using
a quadrature rule obtained by trigonometric interpolation, while the smooth
contribution containing $K_2$ is evaluated with the periodic trapezoidal
rule. For analytic boundaries and data, the corresponding quadrature errors
decrease exponentially with the number of discretization points.

For piecewise-smooth boundaries, Kress quadrature is combined with a graded
parametrization that clusters nodes near true geometric corners. The grading
strength and boundary resolution must be increased together: excessive
grading at fixed \(N\) can place nodes too close to one another, whereas
insufficient grading leaves the singular corner behavior unresolved. The
geometry representation, grading maps, and practical convergence behavior are
described in Appendix~\ref{sec:geometry}.

For multiply connected geometries the singular quadrature is required only
for the same closed boundary components. Interactions between distinct,
non-touching components are smooth and are evaluated by trapezoidal
quadrature. At fixed points per wavelength the number of boundary degrees of freedom
grows approximately linearly with $k$~\cite{backer_BIM,barnett_NtD},
\begin{equation}
    N=O(k).
\end{equation}
A dense matrix factorization therefore costs $O(N^3)=O(k^3)$. In a direct
sweep this cost is incurred repeatedly, including at wavenumbers which are not the wanted solutions. Direct sweeps are consequently valuable for inspection,
validation, and narrow searches, but become extremely inefficient when many
eigenvalues are required. This motivates the accelerated eigensolvers
described next.

\section{Accelerated solvers}
\label{sec:accelerated_solvers}

The accelerated solvers replace a dense scan of individual trial
wavenumbers by a local spectral problem that returns several eigenvalues at
once. \texttt{QuantumBilliards.jl} provides four such methods:
Vergini--Saraceno (VS) \cite{VerSar1995,barnett_dissipation}, the expanded boundary integral method (EBIM) \cite{VebProRob2007,dietz_neutrino_expanded}, Beyn's
contour-integral method \cite{beyn_contour}, and CORK \cite{CORK, CORK_practical}. VS operates on a Helmholtz basis, while EBIM, Beyn, and CORK operate directly on the nonlinear boundary-integral problem of Eq.~\eqref{eq:bim_nep}. Their numerical details differ, but they share the same high-level interface: a spectral interval is divided into windows, the local problem is solved in each window, and candidate
eigenpairs are filtered and combined into the final spectrum.

The main ideas and user-facing parameters are summarized below. Numerical
rank reduction, window selection, contour convergence, polynomial
approximation, and other implementation details are collected in
Appendices~\ref{sec:vs_details}--\ref{sec:cork_polynomial}.

\subsection{Vergini--Saraceno scaling method}
\label{sec:vs}

The Vergini--Saraceno scaling method
~\cite{VerSar1995,Barnett2000,barnett_NtD,BarBet2007} converts the basis
problem of Sec.~\ref{sec:basis_formulation} into a local generalized
eigenvalue problem around a reference wavenumber $k_0$. The idea comes from a Taylor expansion of a weighted version of the tension norm \eqref{eq:tension_norm} around this reference wavenumber. Let $G$ denote
the boundary collocation matrix and $\partial_kG$ its derivative,
\begin{equation}
    G_{ij}
    =
    \phi_j(\mathbf{x}_i;k_0),
    \qquad
    (\partial_kG)_{ij}
    =
    \left.
    \frac{\partial\phi_j(\mathbf{x}_i;k)}
         {\partial k}
    \right|_{k=k_0}.
    \label{eq:vs_G_dG}
\end{equation}
Using the scaling weights
\begin{equation}
    w_i
    =
    \frac{\Delta s_i}
         {\mathbf{x}_i\cdot\mathbf{n}_i},
    \qquad
    W=\operatorname{diag}(w_i),
    \label{eq:vs_weights}
\end{equation}
incorporates the natural boundary measure associated with a uniform dilation
of the billiard. This weighting is closely related to the
quasiorthogonality of boundary functions underlying the scaling
method~\cite{VerSar1995,barnett_dissipation}. In particular, the normal
derivatives of nearby eigenstates are approximately orthogonal under the
dilation-weighted boundary inner product involving
$\mathbf{x}\cdot\mathbf{n}$. This quasiorthogonality is what allows a
cluster of eigenstates near the reference wavenumber to be extracted from a
single generalized eigenvalue problem. Formally with a single expansion center is this method only applicable to star shaped billiards \cite{Barnett2000}. One then forms
\begin{equation}
    F=G^\dagger WG,
    \qquad
    F_k
    =
    (\partial_kG)^\dagger WG
    +
    G^\dagger W(\partial_kG),
    \label{eq:vs_F_Fk}
\end{equation}
and solves
\begin{equation}
    F_k\mathbf c=\mu F\mathbf c.
    \label{eq:vs_gevp}
\end{equation}
Each admissible generalized eigenvalue gives an approximation \cite{Barnett2000}
\begin{equation}
    k
    =
    k_0-\frac{2}{\mu}
    +\frac{2}{k_0\mu^2}
    +O(\mu^{-3}).
    \label{eq:vs_result}
\end{equation}
Thus a single generalized eigenvalue problem returns several eigenvalues near
$k_0$, rather than testing them individually, which would destroy a large part of the method's efficiency.

In addition to the basis dimension \eqref{eq:basis_size} and number of collocation points \eqref{eq:bim_size}, we may also control the size of the spectral window around each expansion center. The parameter $dk$ is the width of the local spectral window used by the scaling method (typically one should decrease it with increasing $k$). No strict error bounds for the eigenvalues are proven for this method \cite{Barnett2000}. In practice the error is approximated by the tension $\Delta k_n\;\le\;C\,\tau_n(k)$ The window size can be adjusted to exclude solutions with tensions larger than some set value. Convergence should be checked by increasing the basis and boundary resolutions. A good criterion is making sure a tiny numerical nullspace forms, signifying that the basis is complete and then removing it for a truncated basis. The corresponding window selection choice is described in Appendix~\ref{sec:vs_details}.

\subsection{Expanded boundary integral method}
\label{sec:ebim}

The expanded boundary integral method (EBIM)
~\cite{VebProRob2007,dietz_neutrino_expanded} applies the same local spectral
idea directly to the nonlinear boundary-integral equation \eqref{eq:bim_nep}. Around a reference wavenumber $k_0$, writing $k=k_0+\varepsilon$, the Fredholm matrix is expanded to $2$nd order
\begin{equation}
    A(k_0+\varepsilon)
    =
    A_0+\varepsilon A_1
    +\frac{\varepsilon^2}{2}A_2
    +O(\varepsilon^3),
    \label{eq:ebim_taylor}
\end{equation}
where $A_j$ denotes the corresponding derivative of $A$ at $k_0$.
To first order, the spectral shifts follow from
\begin{equation}
    A_0\mathbf v
    =
    \lambda A_1\mathbf v,
    \qquad
    \varepsilon_1=-\lambda.
    \label{eq:ebim_gevp}
\end{equation}
For a corresponding left generalized eigenvector $\mathbf u$, the
second-order correction is
\begin{equation}
    \varepsilon_2
    =
    -\frac{\varepsilon_1^2}{2}
    \frac{\mathbf u^\dagger A_2\mathbf v}
         {\mathbf u^\dagger A_1\mathbf v},
    \label{eq:ebim_second_order}
\end{equation}
giving
\begin{equation}
    k_{\rm EBIM}
    =
    k_0+\varepsilon_1+\varepsilon_2.
    \label{eq:ebim_result}
\end{equation}
The important practical distinction from a sweep is that the Fredholm matrices
and their derivatives are constructed only at the expansion centers, after
which several nearby eigenvalues are obtained from the generalized eigenvalue
problem. The principal additional control is the spectral window around each
expansion center, specified by $dk$. Since the expansion is carried
through second order, the first neglected term is $O(\varepsilon^3)$, where
$\varepsilon=k-k_0$. Smaller spectral windows therefore improve the accuracy
at the expense of requiring more expansion centers. Matrix and derivative
evaluations can additionally be accelerated using Taylor approximations to the Fredholm matrix (see Appendix~\ref{sec:cork_polynomial}). The Taylor approximation is controlled by additional numerical parameters. More details are
given in Appendix~\ref{sec:ebim_details}.

\subsection{Beyn's contour-integral method}
\label{sec:beyn}

Beyn's method~\cite{beyn_contour,sakurai_contour} treats
Eq.~\eqref{eq:bim_nep} directly as a nonlinear eigenvalue problem. Rather
than approximating $A(k)$ locally, it extracts all eigenvalues enclosed by
a (in our implementation circular) contour $\Gamma = \{k_0+R\exp(\mathrm{im}\phi) | \phi \in (0,2\pi)\}$ in the complex $k$-plane. The idea is that $A(z)^{-1}$ is a meromorphic function of $z$ with poles at the nonlinear eigenvalues. Integrating it around a contour picks up the residues of the poles inside. Those residues contain the eigenvalues and eigenvectors, and the moment matrices $M_0$ and $M_1$ encode them in a form that reduces to an ordinary eigenvalue problem.
For a probing matrix $V\in\mathbb C^{N\times L}$, the first two contour
moments are
\begin{equation}
    M_0
    =
    \frac{1}{2\pi i}
    \oint_\Gamma A(z)^{-1}V\,dz,
    \qquad
    M_1
    =
    \frac{1}{2\pi i}
    \oint_\Gamma zA(z)^{-1}V\,dz.
    \label{eq:beyn_moments}
\end{equation}
If the numerical rank of $M_0$ is $r$, its truncated singular-value
decomposition is
\begin{equation}
    M_0
    =
    U_r\Sigma_rW_r^\dagger.
    \label{eq:beyn_svd}
\end{equation}
The original $N$-dimensional nonlinear problem is then reduced to
\begin{equation}
    B
    =
    U_r^\dagger M_1W_r\Sigma_r^{-1},
    \label{eq:beyn_reduced}
\end{equation}
whose eigenvalues approximate the nonlinear eigenvalues enclosed by
$\Gamma$. The associated eigenvectors provide approximations to the
boundary densities.

\begin{figure}[h!]
    \centering
    \includegraphics[width=\linewidth]{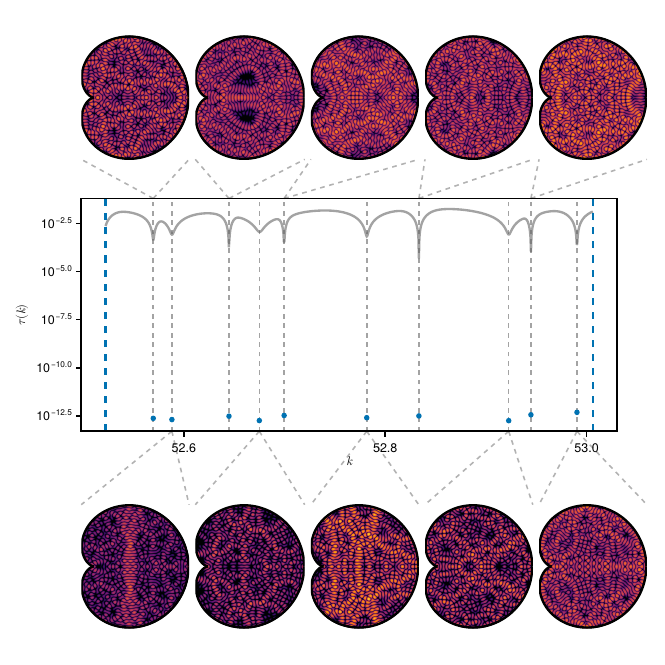}
    \caption{Poles of the inverse of the Fredholm matrix $k$ as found via Beyn's method. The $10$ low-lying states were obtained with a single \texttt{solve} call. The blue points show the residual $\tau(k) = \frac{||A(k)v(k)||}{||v(k)||}$ of each pole, where $v(k)$ is right singular vector of the smallest singular value (all $\tau(k)$ must be smaller than \texttt{res\_tol}). In  gray is overlayed a traditional sweep over the $k$ interval to show that no poles have been missed. The solver for the underlying quadrature was \texttt{DoubleLayerPotentialSolver(10,LimaconBilliard(1.0))}. The contour was discretized using \texttt{nq=40} trapezoidal nodes and all residuals are below \texttt{res\_tol=1e-10}. The wavefunction color-plots are shown in the logarithmic scale.}
    \label{fig:beyn_run}
\end{figure}

For a requested interval $[k_0-R,k_0+R]$, the library automatically constructs a
sequence of circular contours using the Weyl estimate for the local density
of states. The principal numerical parameters are the maximum contour radius, the number of contour quadrature nodes, and the
initial probing rank $r$. The latter is increased automatically if the
numerical rank of the moment matrix approaches the available probing
dimension. Aditional parameters set the numerical-rank cutoff in
Eq.~\eqref{eq:beyn_svd} and  the residual threshold
used to validate the resulting nonlinear eigenvalue and eigenvector pairs. An example
of a run is shown in Fig.\ref{fig:beyn_run}, producing 10 states with a single solution call.

The method requires evaluation of Hankel functions with complex arguments, which is notoriously expensive \cite{AmosHankel}. Construction of the Fredholm matrices on the contour can be accelerated by
expanding them around the contour center $k_0$. By default, the solver uses the
analytic Taylor--Horner backend (see Appendix~\ref{sec:cork_polynomial}) Chebyshev interpolation is
also available. These accelerations affect the evaluation of
$A(k)$, while the contour-moment formulation itself remains unchanged.

The choice of contour discretization, probing rank, and SVD cutoff is
discussed in detail in Appendix~\ref{sec:beyn_details}. For the
calculations considered here, approximately $40$ contour nodes are
generally sufficient, while the appropriate SVD and residual tolerances
depend on the desired accuracy and the underlying boundary discretization.

\subsection{Compact rational Krylov method}
\label{sec:cork}

The \textit{compact rational Krylov solver} (CORK) also solves the nonlinear boundary-integral problem
\begin{equation}
    A(k)\mathbf u=0,
    \label{eq:cork_nep}
\end{equation}
but replaces the dependence of $A(k)$ on $k$ within each spectral window
by a Chebyshev polynomial approximation \cite{CORK_practical,CORK}. Consider a window centered at $k_0$ with requested half-width $\Delta$. The approximation is constructed on a slightly larger guarded interval,
\begin{equation}
    \Delta_p=(1+g)\Delta,
    \label{eq:cork_guard}
\end{equation}
where $g>0$ is a small guard parameter and the rescaled coordinate
\begin{equation}
    t
    =
    \frac{k-k_0}{\Delta_p}
    \label{eq:cork_mapping}
\end{equation}
maps this interval to $[-1,1]$. The Fredholm matrix is approximated by
\begin{equation}
    A(k_0+\Delta_p t)
    \simeq
    P(t)
    =
    \sum_{j=0}^{p}\frac{\Delta_p^j}{j!}A^{(j)}(k_0)t^j
    =
    \sum_{j=0}^{p}B_jT_j(t),
    \label{eq:cork_chebyshev}
\end{equation}
where $T_j$ is the Chebyshev polynomial of degree $j$. $B_j$ are therefore internally constructed from the matrices corresponding to the derivatives of the Fredholm matrix $A^{(j)}(k_0)=\frac{\partial^j A(k_0)}{\partial k^j}$ and the polynomial basis transformation $t_j \rightarrow T_j(t)$ (see Appendix~\ref{sec:cork_polynomial} for more details). The nonlinear eigenproblem \eqref{eq:cork_nep} is thereby replaced locally by the polynomial eigenvalue problem
\begin{equation}
    P(t)\mathbf u=0.
    \label{eq:cork_pep}
\end{equation}
A direct construction of a degree-$p$ matrix polynomial would introduce a
problem of dimension $O(pN)$. The compact Krylov formulation exploits the structured formulation induced by the three-term Chebyshev recurrence, representing the Krylov basis in factored form so that the large vectors are stored in the original $N$-dimensional space~\cite{CORK_practical,CORK}. This substantially reduces the storage and orthogonalization cost of the linearized problem.

The implementation uses a shift--invert formulation centered on the spectral
window. If $\mu$ denotes the corresponding transformed eigenvalue, then
\begin{equation}
    t=\frac{1}{\mu},
    \qquad
    k=k_0+\frac{\Delta_p}{\mu}.
    \label{eq:cork_recovery}
\end{equation}
For calculations over a large spectral interval, the library automatically
divides the requested range into smaller windows using the Weyl estimate.
The principal CORK-specific parameters are the target number of levels per
window, the maximum window half-width, the
Chebyshev degree $p$, and the guard factor $g$. The first
two control the automatic subdivision of the spectrum, while $p$
controls the accuracy with which $A(k)$ is represented over each window.
The guard parameter $g$ enlarges the polynomial approximation interval beyond the
requested spectral window, reducing sensitivity to its edges. Increasing
$p$ improves the polynomial approximation for a fixed window, whereas
reducing the window width decreases the polynomial degree required for a
given accuracy. The underlying BIE discretization resolution ultimately limits the attainable
accuracy.

The compact Krylov dimension is increased automatically until the accepted
spectrum is stable and the roots near the spectral-window boundaries have
converged. A parameter sets the residual tolerance used
to accept candidate eigenpairs. The Chebyshev recurrence, compact
formulation, convergence criteria, and treatment of the final spectral
window are described in Appendix~\ref{sec:cork_polynomial}. 

\section{Representations of eigenstates}
\label{sec:wav_hus}

Computing the eigenvalues is only the first stage of many
calculations. \texttt{QuantumBilliards.jl} gives a common interface from which the boundary function, wavefunction, and Poincar\'e--Husimi
function can be constructed.

\subsection{Boundary function and layer density}
\label{sec:boundary_function_layer}

The boundary function is the outward normal derivative of the
Dirichlet eigenfunction~\cite{backer_boundary,Backer2003},
\begin{equation}
    u(s)
    =
    \partial_n\psi(\mathbf r(s)),
    \label{eq:u_def}
\end{equation}
where $s$ denotes arclength. It provides a natural boundary representation
of a Dirichlet eigenstate and is also the input used for the
Poincar\'e--Husimi representation below.

For a basis state
\begin{equation}
    \psi(\mathbf r)
    =
    \sum_\alpha c_\alpha\phi_\alpha(\mathbf r),
    \label{eq:wav_basis}
\end{equation}
the boundary function follows directly from
\begin{equation}
    u(s_i)
    =
    \mathbf n(s_i)\cdot
    \sum_\alpha
    c_\alpha\nabla\phi_\alpha(\mathbf r_i).
    \label{eq:u_from_basis}
\end{equation}
Boundary-integral solvers return a layer density $\mu$ (smallest right singular vector of \eqref{eq:bim_nep} at an eigenvalue of \eqref{eq:helmholtz}), which is distinct from the physical boundary function $u=\partial_n\psi$. For the DLP formulation, the physical boundary function is recovered from the corresponding left singular vector $\ell$ as $\ell^{\dagger} A(k)=0$. With $W=\operatorname{diag}(ds_i)$, the weighted transpose is (where $ds_i$ is the arclength quadrature weights at point $i$ on the boundary)
\begin{equation}
A_{\mathrm{adj}}=W^{-1}A^T W.
\end{equation}
Since $A^\dagger\ell = 0$, it follows that
\begin{equation}
A_{\mathrm{adj}}\bigl(W^{-1}\overline{\ell}\bigr) = A_{\mathrm{adj}}\bigl(u_{\mathrm{raw}}) =0,
\end{equation}
where $\overline{\ell}$ is the complex conjugate of $\ell$ and $u_{\mathrm{raw}}=W^{-1}\overline{\ell}$ is the unnormalized boundary function. Its overall amplitude is subsequently fixed using the Rellich identity \eqref{eq:rellich}.

For CFIE, $\mu$ is the density of the combined potential. The physical boundary function can be obtained by taking the normal derivative of the combined potential. The hypersingular contribution is handled easier using Maue's identity~\cite{maue1949,kress_book}, giving
\begin{align}
    u(\mathbf x)
    ={}&
    -\partial_s
    \int_\Gamma
    G_k(\mathbf x,\mathbf y)
    \,\partial_{s_y}\mu(\mathbf y)\,ds(\mathbf y)
    \nonumber\\
    &-
    k^2\mathbf n(\mathbf x)\cdot
    \int_\Gamma
    G_k(\mathbf x,\mathbf y)
    \,\mathbf n(\mathbf y)\mu(\mathbf y)\,ds(\mathbf y)
    \nonumber\\
    &-
    ik\left[
    \frac12\mu(\mathbf x)
    +
    \int_\Gamma
    \frac{\partial G_k(\mathbf x,\mathbf y)}
         {\partial n_{\mathbf x}}
    \mu(\mathbf y)\,ds_{\mathbf y}
    \right],
    \label{eq:u_normal_der_cfie}
\end{align}
where $\partial_s$ is the arclength derivative on the boundary. The same reconstruction is applied on each connected boundary component, with the integral operators retaining the coupling between components. The reason for adding this additional complexity is due to the fact that most BIM accelerated solvers give the layer density "for free" with the eigenvalues, therefore making these conversions useful.

For all formulations, the overall normalization is fixed without performing
an interior area integral by using the Rellich identity~\cite{rellich1940},
\begin{equation}
    \|\psi\|_{L^2(\Omega)}^2
    =
    \frac{1}{2k^2}
    \int_{\partial\Omega}
    (\mathbf r\cdot\mathbf n)
    |u(\mathbf r)|^2\,ds.
    \label{eq:rellich_boundary}
\end{equation}
Thus, irrespective of whether the eigenstate originates from a basis or
boundary-integral formulation, the library provides the same normalized
physical boundary function $u(s)$ for subsequent calculations.

\subsection{Wavefunction}
\label{sec:wav}

Once an eigenstate has been obtained, its interior wavefunction can be
evaluated on arbitrary points in the billiard. For basis states this is
simply the expansion in Eq.~\eqref{eq:wav_basis}.

For states constructed via the DLP approach \eqref{eq:dlp_operator} the wavefunction may instead be reconstructed directly from the layer density $\mu$ \cite{colton_kress},
\begin{equation}
    \psi(\mathbf x)=
    \int_\Gamma
    \frac{\partial G_k(\mathbf x,\mathbf y)}
         {\partial n(\mathbf y)}
    \mu(\mathbf y)\,ds(\mathbf y).
    \label{eq:wav_from_dlp}
\end{equation}
Since the DLP density is directly related to $u=\partial_n\psi$, the
normalized physical boundary function provides the equivalent real
representation \cite{backer_BIM}
\begin{equation}
    \psi(\mathbf x)=
    -\frac14
    \int_{\Gamma}
    Y_0\!\left(k|\mathbf x-\mathbf y|\right)
    u(\mathbf y)\,ds(\mathbf y),
    \label{eq:wav_from_u}
\end{equation}
where $Y_0$ is the Bessel function of the second kind. If $u$ has been
normalized using Eq.~\eqref{eq:rellich}, this representation directly yields
an $L^2$-normalized eigenfunction.

For states constructed via the CFIE approach \eqref{eq:cfie_operator} the layer density $\mu$ is distinct from $u$, and the interior wavefunction is reconstructed from the combined potential \cite{colton_kress},
\begin{equation}
    \psi(\mathbf x)=
    -\int_\Gamma
    \frac{\partial G_k(\mathbf x,\mathbf y)}
         {\partial n(\mathbf y)}
    \mu(\mathbf y)\,ds(\mathbf y)
    -ik\int_\Gamma
    G_k(\mathbf x,\mathbf y)\mu(\mathbf y)\,ds(\mathbf y).
    \label{eq:wav_from_cfie}
\end{equation}
Evaluation of Hankel functions dominates runtime costs when a
wavefunction is reconstructed on a dense spatial grid. The implementation
therefore provides both direct kernel evaluation and piecewise Chebyshev
interpolation of the radial functions.

\subsection{Poincar\'e--Husimi function}
\label{sec:hus}

The Poincar\'e--Husimi (PH) function provides a phase-space representation of
an eigenstate through its physical boundary function $u(s)$
~\cite{Backer2003,TV1995}. For the boundary coordinate $q$ and tangential
momentum $p\in[-1,1]$, the PH amplitude is defined as
\begin{equation}
    h(q,p)
    =
    \int_0^L u(s)c_{q,p}(s)\,ds,
    \label{eq:hus_def}
\end{equation}
where
\begin{equation}
    c_{q,p}(s)=
    \sum_{\ell\in\mathbb Z}
    \exp\!\left[-\frac{k}{2}(s-q+\ell L)^2\right]
    \exp\!\left[-ikp(s-q+\ell L)\right]
    \label{eq:tuall_voros_coherent}
\end{equation}
is a periodized Gaussian coherent state. Its spatial width
$\sigma=1/\sqrt{k}$ sets the phase-space resolution on the natural
semiclassical scale. The corresponding PH density is
\begin{equation}
    H(q,p)=|h(q,p)|^2,
    \label{eq:hus_sq}
\end{equation}
and is normalized numerically to sum to unity.

For boundary functions that are real up to a global phase, the PH density
satisfies
\begin{equation}
    H(q,p)=H(q,-p),
    \label{eq:hus_1d_irrep_sym}
\end{equation}
so only $p\geq0$ needs to be evaluated explicitly. For complex states the
full interval $p\in[-1,1]$ is evaluated.

\section{Code architecture and examples}
\label{sec:code}
The library is composed of the main package \texttt{QuantumBilliards.jl} and the helper package \texttt{BilliardGeometry.jl} that handles all geometric operations such as evaluating the boundary points and quadratures and houses the geometric definitions of the billiards. The user may install the packages with the package manager:
\begin{lstlisting}[language=Julia]
using Pkg

Pkg.add("QuantumBilliards")
Pkg.add("BilliardGeometry")
\end{lstlisting}
This section will explain the core principles of using the package. For full details we refer the user to the documentation pages.  

\begin{figure}[h!]
    \centering
    \includegraphics[width=1.0\linewidth]{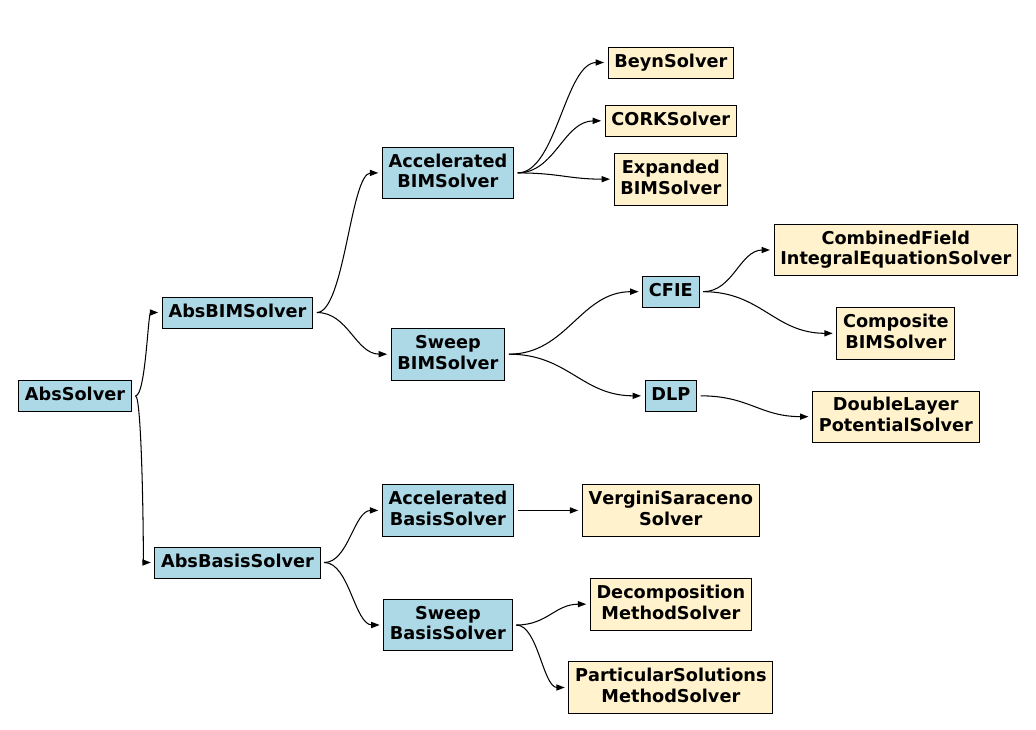}
    \caption{Type diagram of the solver types in the library. Abstract types are shown in blue and concrete types in yellow. The arrows connect each super-type to its subtypes.}
    \label{fig:Diagram}
\end{figure}

We leverage Julia's multiple dispatch system to construct mathematically meaningful abstractions and organize them in a type hierarchy and then dispatch functions on these types. For instances the solvers are organized in the type hierarchy shown in Fig. \ref{fig:Diagram}. This allows for a separation for the API layer from the specific implementation of the solvers and a very modularized approach. The benefits are twofold. Firstly, we may implement extremely specialized functions dispatched on the lower levels of the hierarchy most often for the concrete types themselves that may also easily be replaced if a newer more optimized implementation becomes available. Secondly, adding a new concrete type with entirely new functionality is easily possible because much of the the infrastructure is already available, since it is dispatched on the upper levels of the hierarchy. Thus we only need to implement the most specialized parts for the new type and match the output to the already existing API. For example if we wish to add a new basis for the basis solvers, we implement a new concrete basis type and functions that evaluate it and the gradients, but we do not need to implement any changes to the basis solvers or spectral computation functions. Each basis solver will be able to use the new basis as long as the output types of the new functions match or are subtypes of the already existing interface. The whole library is designed in this modular fashion making it very extensible and tunable to individual needs. The authors intend to continue to evolve the code with new additions, like hyperbolic and magnetic billiards (see Appendix \ref{sec:appendix_future}).

\subsection{Example workflow}
\label{sec:workflow}
\begin{figure}[h!]
    \includegraphics[width=\linewidth]{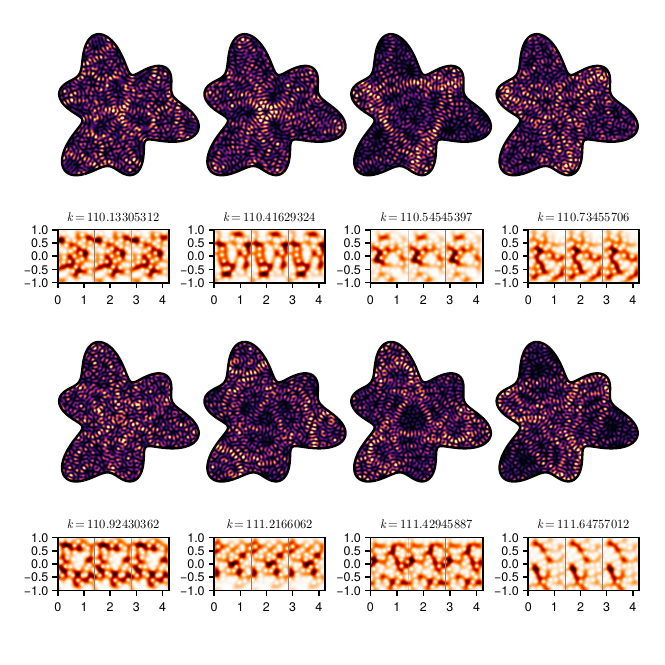}
    \caption{Color-plots of consecutive eigenstates of the $C_3$ billiard in the wavefunction and Poincaré–Husimi representation. The Poincaré–Husimi function is a projection on the classical phase space coordinates $(q,p)$}
    \label{fig:c3_states}
\end{figure}

A spectral computation consists of the following workflow: Define the billiard and its parameters, define the solver and set the numerical parameters, in case of basis solvers define the basis, call the \texttt{compute\_spectrum} for the desired wavenumber or consecutive state number interval. The \texttt{compute\_spectrum} function returns a \texttt{SpectralData} object that contains the computed wavenumbers, tensions as well as optionally the computed eigenstates and is all that is needed to construct the wavefunctions, boundary functions and Husimi functions. The following script computes 8 states starting from the 250-th of a billiard with a $C_3$ rotational symmetry:

\begin{lstlisting}[language=Julia]
using QuantumBilliards, BillardGeometry

c3_billiard = C3Billiard(0.2)

c3_kernel = DoubleLayerPotentialSolver(
    8.0,
    c3_billiard;
    grading = SmoothPeriodicGrading(),
    sym_characters = (1,),
)

c3_beyn = BeynSolver(
            c3_kernel;
            eigenvectors = true
        )
n_states = 8
start_state = 250
k_min, k_max = k_range_for_states(c3_billiard, start_state, start_state + n_states - 1)
c3_spect = compute_spectrum(c3_beyn, c3_billiard, k_min, k_max; multithreaded=true)
\end{lstlisting}
Here we selected the collocation point scaling parameter $b_{\partial\mathcal{B}}=8.0$ (from Eq. \eqref{eq:bim_size}), with no corner grading since the billiard boundary is smooth. The states are computed with Beyn's method and the double layer potential kernel. The symmetry sector is selected with the \texttt{sym\_characters} parameter. In this case the doublet states. The returned \texttt{SpectralData} object also contains all the information about the billiard and solver that produced it and is index-able and slice-able. Thus we may compute the wavefunction of the first state by simply calling
\begin{lstlisting}[language=Julia]
Psi, x_grid, y_grid = wavefunction(c3_spect[1]; b= 5.0, inside_only = true)
\end{lstlisting}
This will automatically construct the appropriate grid with 5 grid-points per wavelength and only evaluate the wavefunction inside the billiard. To save computation costs of grid constructions etc. we may also call it on the \texttt{SpectralData} object directly without selecting an index. This will compute the wavefunctions for all the contained states and return a vector of arrays. We provide similar functions for computing boundary functions and Poincaré–Husimi functions
\begin{lstlisting}[language=Julia]
u, s = boundary_function(c3_spect[1])
h, p, q = husimi_function(c3_spect[1]; full_p=true)
\end{lstlisting}
Additionally, a visualization package \texttt{QBPlotting.jl} using the \texttt{Makie} plotting library is available for users that wish to use it. The state visualizations in this paper were produced with this package. We show the 8 eigenstates from the above example in Fig. \ref{fig:c3_states}. We may observe this is exactly the case where the Poincaré–Husimi are not symmetric with respect to $p$. 

\subsection{Billiard showcase}
\label{sec:showcase}

\begin{figure}[h!]
    \includegraphics[width=\linewidth]{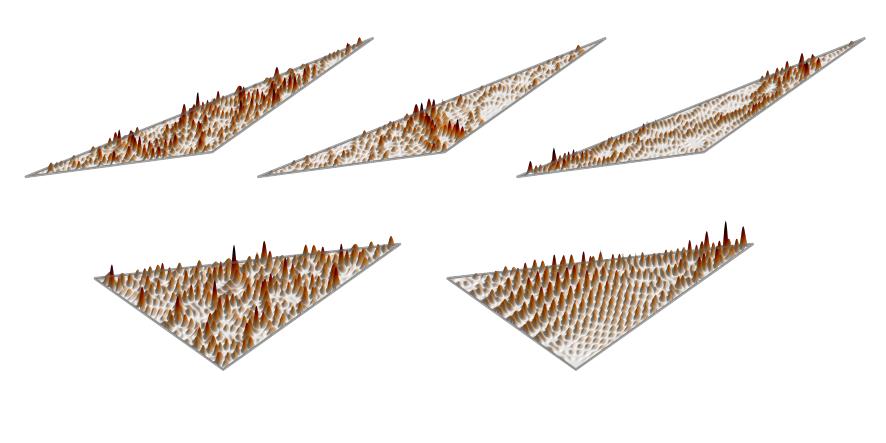}
    \caption{Representative states of triangular billiards. Top row irrational triangle with largest angle \(\gamma = \frac{\pi(\sqrt{5}+1)}{4}\) and ratio between the other angles \(\chi = \frac{1+\sqrt{2}}{2}\). The wavenumbers are $k=80.74998923$, $80.85312668$, $81.47525021$. Bottom row Veech right-triangle \(\gamma = \frac{\pi}{2}\) \(\chi = \frac{3}{2}\). The wavenumbers are $k=100.70678045$, $100.84760162$} 
    \label{fig:showcase_triangles}
\end{figure}

\begin{figure}[h!]
    \includegraphics[width=\linewidth]{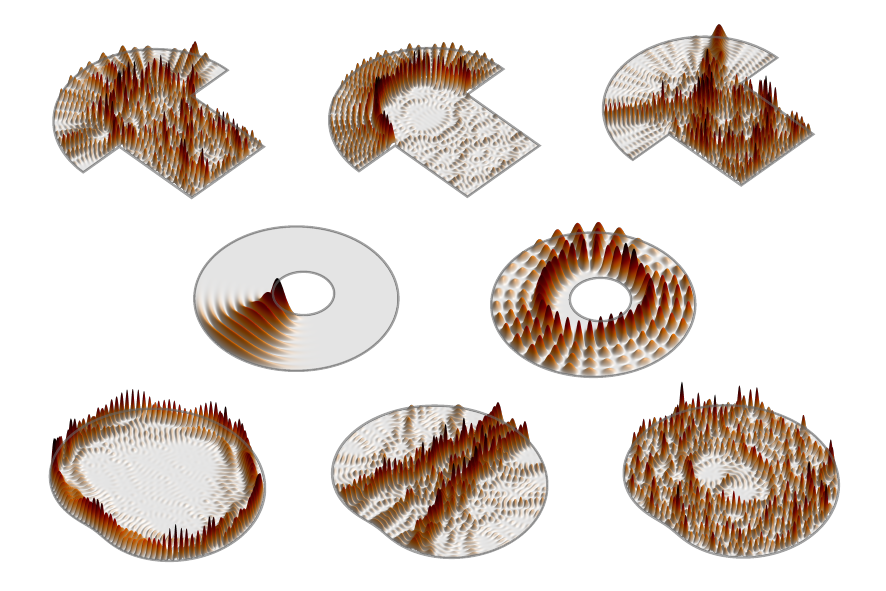}
    \caption{Representative states of billiards with divided phase space. Top row mushroom with circle half width 0.5 and stem height 1.0. The wavenumbers are $k=71.51329513$, $71.62799215$, $72.62405796$. Middle row annular billiard with inner circle radius 0.3 shifted to the right by 0.1 and outer circle radius 1.0. The wavenumbers are $k=31.73927904$, $31.75281169$. Bottom row Robnik billiard with parameter 0.6. The wavenumbers are $k=59.96125494$, $60.45599159$, $60.50673403$.} 
    \label{fig:showcase_mixed}
\end{figure}

\begin{figure}[h!]
    \includegraphics[width=\linewidth]{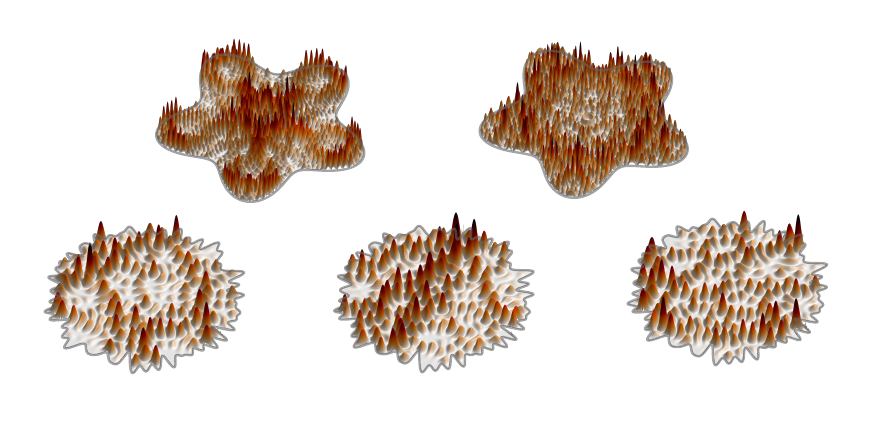}
    \caption{Representative states of the $C_5$ star billiard with parameter value 0.2 (top row). The wavenumbers are $k=71.45070797$, $71.974751$. Ripple billiard \eqref{eq:geometry_fourier} with 100 coefficients (bottom row). The wavenumbers are $k=33.46743963$, $33.50953131$, $33.57496036$} 
    \label{fig:showcase_star}
\end{figure}

The library implements a wide range of billiard geometries as standard Julia objects. In this section we show some representative examples of wavefunctions. In Fig. \ref{fig:showcase_triangles} we show states of one highly irrational triangle (all angles incommensurate with $\pi$) and a Veech right-triangle. Even though the irrational triangles spectrum follows GOE random-matrix spectral statistics \cite{lozej2022triangles} we still observe some scarring. The Veech triangle follows intermediate spectral statistics \cite{lozej_intermediate_2024} and thus some states appear similar to chaotic states and some close to regular. In Fig. \ref{fig:showcase_mixed} we show states of three billiards with divided phase space. The top row shows the mushroom billiard where the first state is a typical chaotic state, the second one, a typical mixed-type state at the edge of the island of stability permeating into the chaotic component, and the third a heavily scarred bouncing-ball like state in the chaotic component. The second row shows the annular billiard, where we used the \texttt{CompositeBIMSolver} with two \texttt{CombinedFieldIntegralEquationSolver} kernels. We see two states form separate regular components, the first one on the period-2 island of stability and the second on a period-3 island very close to a circle eigenstate. The bottom row
shows the Robnik billiard with the parameter value 0.6 which is already non convex. The first state is heavily scarred by boundary glancing orbits, the second is a regular state around an island of stability and the third a typical chaotic state that in this case is not uniform since the chaotic component has a complicated structure because of the islands of stability. In Fig. \ref{fig:showcase_star} we show two representative states of the $C_5$ symmetric star billiard (top row) and three representative states of a random ripple billiard. The ripple billiard is generated by randomly selecting 100 (small compared to 1) expansion coefficients of a Fourier series for the radial function in polar coordinates given by Eq. \eqref{eq:geometry_fourier} in Appendix \ref{sec:geometry}. 

\subsection{Benchmarks}
\label{sec:benchmarks}

\begin{figure}[h!]
    \centering
    \begin{subfigure}[t]{0.49\linewidth}
        \centering
        \includegraphics[width=\linewidth]{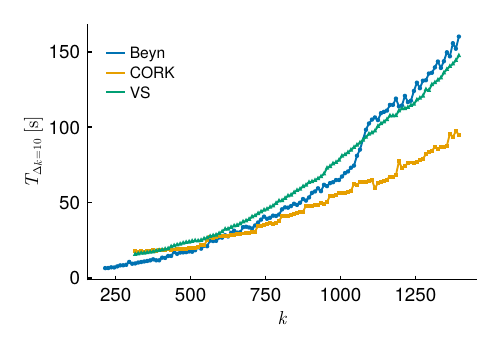}
        \caption{}
        \label{fig:runtime_vs_k}
    \end{subfigure}
    \hfill
    \begin{subfigure}[t]{0.49\linewidth}
        \centering
        \includegraphics[width=\linewidth]{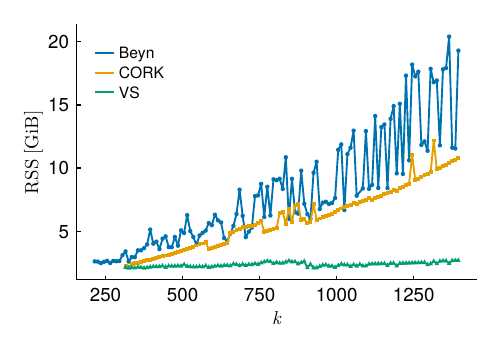}
        \caption{}
        \label{fig:rss_vs_k}
    \end{subfigure}
    \caption{Computational scaling of the accelerated eigensolvers with
    wavenumber. The geometry is the same as used in Table~\ref{tab:large_run}. (a) Each point corresponds to the time needed to make an independent call to
    \texttt{compute\_spectrum} over an interval of width $\Delta k=10$ (labeled as $T_{\Delta k=10}[s]$). (b) Peak resident-set size (RSS). Because every point is obtained from a separate calculation, solver setup
    and preparation are repeated for every interval. The measurements therefore
    characterize both the local scaling with wavenumber and the cost of repeatedly
    initializing short calculations, rather than the optimal throughput of a
    single large-scale calculation (see Table.\ref{tab:large_run}). All benchmarks were performed using Julia
    1.12.7 with MKL.jl on an AMD Ryzen Threadripper PRO 7985WX processor
    (64 physical cores, 128 hardware threads), using 64 Julia threads and
    64 BLAS \cite{BLAS_ref} threads.}
    \label{fig:solver_benchmarks}
\end{figure}

The computational cost of the accelerated solvers depends not only on the
wavenumber, but also on how a large spectral calculation is divided into
individual calls. To expose the local scaling with $k$,
Fig.~\ref{fig:solver_benchmarks} shows the runtime and peak memory consumption
of independent calculations over consecutive intervals of width
$\Delta k=10$. Since each point corresponds to a separate invocation of
\texttt{compute\_spectrum}, solver initialization and preparation are repeated
for every interval.

\begin{figure}[h!]
    \centering
    \includegraphics[width=\linewidth]{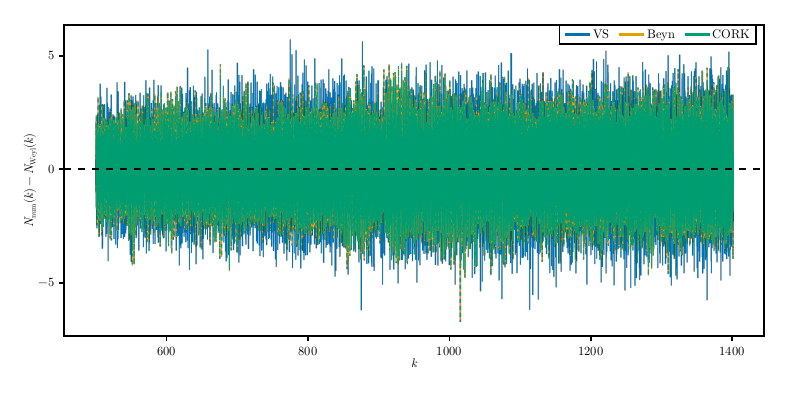}
    \caption{Weyl-law completeness test for the Prosen billiard, showing the difference between the numerical spectral staircase $N_{\mathrm{num}}(k)$ and the Weyl \cite{Weyl1911} prediction $N_{\mathrm{Weyl}}(k)$. No systematic downward drift is observed as $k$ increases, indicate that the computed spectra are complete; missing eigenvalues would instead produce an accumulating negative drift in $N_{\mathrm{num}}(k)-N_{\mathrm{Weyl}}(k)$.}
    \label{fig:weyl_complete}
\end{figure}

For a production calculation over a large spectral range, these repeated
per-call costs can be avoided by requesting the complete interval in a single
call. The high-level solver can then organize the calculation internally
without repeatedly invoking the complete setup and preparation path for
independent short calculations. To illustrate the resulting end-to-end
performance, Table~\ref{tab:large_run} reports calculations of the complete
spectrum over $500\leq k\leq1400$. All reported spectra are complete, as can be seen from Fig.\ref{fig:weyl_complete}. In contrast to the independent
$\Delta k=10$ calculations of Fig.~\ref{fig:solver_benchmarks}, these
measurements therefore represent the intended use of
\texttt{compute\_spectrum} for large-scale calculations. The code is given in Listing \ref{lst:full_spectra}.
\FloatBarrier
\begin{lstlisting}[
    language=Julia,
    caption={Large-scale spectral computations with the \texttt{VerginiSaracenoSolver}, \texttt{BeynSolver}, and \texttt{CORKSolver}.},
    label={lst:full_spectra}
]
using QuantumBilliards, BilliardGeometry, StaticArrays
try_MKL!()
kmin, kmax, ppw = 500.0, 1400.0, 7.0 # acc plateau reached

billiard = ProsenBilliard(0.04) # for larger a no basis conv.
basis = CornerAdaptedFourierBessel(10, Float64(pi / 2), SVector(0.0, 0.0), 0.0)
solver = VerginiSaracenoSolver(2.5, ppw)
data_vs = compute_spectrum(solver, basis, billiard, kmin, kmax, 0.05)
billiard = ProsenBilliard(0.2) 
dlp = DoubleLayerPotentialSolver(ppw, billiard; sym_characters = (-1, -1))
solver = BeynSolver(dlp; nq = 40, m = 200, Rmax = 0.5, svd_tol = 1e-11, res_tol = 1e-6)
data_beyn = compute_spectrum(solver, billiard, kmin, kmax)
solver = CORKSolver(dlp; nlevels = 100, mstart = 400, Rmax = 0.5, p = 15, guard = 0.05, b = 5)
data_cork = compute_spectrum(solver, billiard, kmin, kmax)
\end{lstlisting}
\FloatBarrier

\begin{table}[h!]
    \centering
    \caption{End-to-end performance for a single
    \texttt{compute\_spectrum} call over $500\leq k\leq1400$.
    For \texttt{VerginiSaracenoSolver}, we use
    \texttt{ProsenBilliard(a)} with \texttt{a=0.04} due to issues with
    basis convergence for stronger deformations, while \texttt{a=0.2}
    is used for the other solvers (therefore level counts differ). To facilitate comparison with the
    basis-based solver, the BIM-based solvers use
    \texttt{XYAxisReflection()} with parity $-1$ under both reflections.
    The throughput is the number of computed eigenvalues divided by the
    total wall time (average). The spectra were generated with their respective solver' default parameters.}
    \label{tab:large_run}
    \renewcommand{\arraystretch}{1.15}
    \begin{tabular}{lrrr}
        \toprule
        Solver & Eigenvalues & Runtime [s] & Eigenvalues/s \\
        \midrule
        \texttt{VerginiSaracenoSolver} & 106\,697 & 2875 & 37.12 \\
        \texttt{BeynSolver}             & 108\,710 & 3338 & 32.56 \\
        \texttt{CORKSolver}             & 108\,710 & 4125 & 26.35 \\
        \bottomrule
    \end{tabular}
\end{table}

The two benchmarks emphasize different aspects of solver performance.
Independent short calculations repeatedly incur preparation costs and are
therefore useful for examining local scaling with $k$, whereas a single
large call amortizes these costs over the complete spectrum. The latter is
the appropriate benchmark for the throughput of a production spectral
calculation. Memory requirements provide an additional distinction between
the methods: the local measurements in Fig.~\ref{fig:rss_vs_k} show
substantially different memory scaling even when the solvers compute the
same spectral intervals. A comparison of their computational complexity and typical use cases is given in Appendix~\ref{sec:summary_accelerated}.

\section{Conclusion}
\label{sec:conclusion}

We presented \texttt{QuantumBilliards.jl}, a Julia framework for the
two-dimensional interior Dirichlet Helmholtz eigenproblem. The library
combines basis-expansion and boundary-integral formulations with direct and
accelerated eigensolvers, including the Vergini--Saraceno, EBIM, Beyn, and
CORK methods. These methods share a common treatment of geometry, discrete
symmetries, spectra, and eigenstates, allowing independent numerical
formulations to be applied to the same billiard problem and their results to
be cross-validated. The framework further provides normalized boundary
functions, interior wavefunctions, and Poincar\'e--Husimi functions,
and supports smooth, cornered, symmetry-reduced, and multiply connected
geometries.

Let us mention the package is the second in the series devoted to the study of quantum chaos. The already published \texttt{SpectralStatistics.jl} package \cite{LozejSpectralStatistics} enables a deep analysis of the spectra by providing computation methods for all the most common spectral statistics and analytical models based on random matrix theory. Additionally, a companion package \texttt{ClassicalBilliards.jl} \cite{LozejClasssicalBilliards} devoted to the classical billiard problem is under development and shares a common interface with the \texttt{BilliardGeometry.jl} package.

\section{Acknowledgements}

MO thanks Prof.\ Dr.\ Marko
Robnik for reading the manuscript and for helpful suggestions, Alex Barnett
for his informative PhD thesis (and elucidating papers) and the development of MPSpack, which provided important inspiration and resolutions for the present numerical framework, and Gregor Veble for his
original \texttt{C++} implementation of EBIM and getting me interested in BIM. ČL thanks Benjamin Batistić for his original Python implementation that served as the first reference, Alex Barnett, Natasha Lepore and Gregor Veble for their highly detailed and informative PhD thesis,
Barbara Dietz, Roland Ketzmerick, Arnd B\"acker, Tomaž Prosen, Giulio Casati, Eugene Bogomolny and Marko Robnik for enlightening discussions and support with a wealth of knowledge of quantum billiard problems as well as Matt Eiles, Frederic Hummel, Aileen Durst, Neethu Abraham, Milena Simi{\'c} for testing the first versions of the package with only basis methods and the Max Planck Society for its hospitality in the first year of development. We acknowledge the use of AI for verifying the derivations in the appendix and assistance in some algorithm optimizations.

\subsection{Funding information}
This work was supported by the Slovenian Research and Innovation Agency
(ARIS) under grants J1-4387 and P1-0306, and made use of the HPC VEGA
supercomputer system under project S24O02-01.

\subsection{Author contributions}
M.O. implemented all solvers and functions that depend on Green functions, the Taylor recurrence schemes and Chebyshev backends, the Kress quadrature scheme, made the project's optimizations and contributed to the geometry package. Č.L. initiated and maintains the project, designed the code architecture, API, CI/CD infrastructure and implemented the geometry package, all basis-dependent solvers and functions, the Poincaré–Husimi function, quadrature schemes belonging to non-singular line integrals and the optional plotting packages. M.O. and Č.L. wrote the manuscript sections corresponding to their respective contributions and jointly reviewed and revised the manuscript.

\bibliography{main}

\section{Appendices}
\appendix

\section{Geometry, boundary discretization, and symmetries}
\label{sec:geometry}

The geometry section is independent of the eigensolvers and provides the common
description from which basis, boundary-integral, and classical calculations
are constructed. It is included as an independent package \texttt{BilliardsGeoumetry.jl}. A billiard is assembled from parametrized boundary curves,
which are grouped into simple or composite domains and, when applicable,
supplemented by a \texttt{SymmetryRegistry} of discrete symmetries.

\subsection*{Boundary curves and domains}

The elementary boundary objects are subtypes of \texttt{AbsCurve}. Currently
implemented examples include straight line segments, circular arcs, and polar
curves. A general polar segment is specified by a radial function
\(r(\phi)\), while \texttt{FourierCoeffPolarSegment} provides the convenient
specialization
\begin{equation}
    r(\phi)
    =
    R+\sum_{n}
    \left[
        a_n\cos(n\phi)+b_n\sin(n\phi)
    \right].
    \label{eq:geometry_fourier}
\end{equation}
This includes, for example, the smooth deformations used for ripple
billiards. Each curve stores its parametrization, orientation, physical
length, boundary condition, and identifiers for the physical domain and
boundary segment to which it belongs.

At the user level, elementary curves can therefore be constructed directly,
for example as
\begin{lstlisting}[language=Julia]
line = LineSegment(p1, p2)
arc = CircleSegment(R, arc_angle, shift_angle, center)
polar = PolarSegment(r_func; R=R)
fourier = FourierCoeffPolarSegment(coef; R=R)
\end{lstlisting}
and combined into domains. Polygonal domains are specified directly by their
vertices,
\begin{lstlisting}[language=Julia]
domain = Polygon([p1, p2, p3, p4], 1)
\end{lstlisting}
with the corresponding line segments, lengths, orientations, and corner
angles constructed automatically. Composite domains provide the analogous
description when the fundamental region is assembled from several elementary
pieces. Different curve segments can carry different boundary roles. In addition to
the physical boundary condition, internal interfaces can be marked as
transparent and boundaries of a fundamental domain can be identified as
symmetry walls. The same geometric representation can therefore describe
both the full physical boundary and a symmetry-reduced fundamental domain.

The stadium billiard is a representative example. Only one quarter of the
geometry is specified explicitly: a quarter-circle is joined to a rectangle,
their common edge is transparent, and the remaining artificial edges are
symmetry walls. The complete billiard is reconstructed by the $D_2$
symmetry group. Its constructor is schematically

\begin{lstlisting}[language=Julia]
function StadiumBilliard(half_width::T) where T<:Real
    symmetries = D2_symmetry()

    circle_dom = CircleWedge(
        one(T), T(pi/2), zero(T), half_width, zero(T), 1;
        bcs = [SpecularReflection(),
               Transparent(2),
               SymmetryWall(3,4)]
    )

    rectangle_dom = Polygon(
        [[half_width,one(T)], [zero(T),one(T)],
         [zero(T),zero(T)], [half_width,zero(T)]], 2;
        bcs = [SpecularReflection(),
               SymmetryWall(1,4),
               SymmetryWall(3,4),
               Transparent(1)]
    )

    return StadiumBilliard{T}(
        CompositeDomain([circle_dom,rectangle_dom]),
        symmetries
    )
end
\end{lstlisting}
Thus the billiard object itself contains the fundamental domain and the
geometric transformations required to reconstruct the complete billiard.

\subsection*{Boundary discretization and corner grading}

For boundary-integral calculations, the physical curves are converted into a
common discrete boundary representation. Each boundary point contains the
coordinates $\mathbf r_i$, outward normal $\mathbf n_i$, physical
arclength $s_i$, and quadrature weight, together with additional geometric
quantities such as tangents and curvature when required by the particular
integral formulation. Connected physical boundary components remain
distinguished throughout this construction.

The principal resolution parameter is the number of points per wavelength,
\texttt{ppw}. The number of boundary points therefore grows approximately
linearly with $k$, maintaining a comparable physical resolution across a
spectral calculation.

The user-facing boundary parametrization is selected through the
\texttt{BoundaryGrading} hierarchy,
\begin{lstlisting}[language=Julia]
SmoothPeriodicGrading()
CornerGrading(kressq = 2, min_t_spacing = 1e-12)
GlobalCornerGrading(kressq = 2, min_t_spacing = 1e-12)
\end{lstlisting}
\texttt{SmoothPeriodicGrading} is the default for smooth closed boundaries
and uses a non-graded trapezoidal quadrature rule. \texttt{CornerGrading} applies
Kress grading to a single parametrized curve containing known corner
singularities, whereas \texttt{GlobalCornerGrading} is intended for
piecewise-smooth composite boundaries. In the latter case, true corners are
detected from tangent discontinuities between adjacent segments; smooth joins
are left ungraded.

For a boundary containing several corners $c_j$, the global construction
acts separately on each interval $[c_j,c_{j+1}]$. Introducing the local
coordinate
\begin{equation}
    u
    =
    \frac{\sigma-c_j}{c_{j+1}-c_j},
    \qquad 0\leq u\leq1,
    \label{eq:kress_local_coordinate}
\end{equation}
the grading function is
\begin{equation}
    g_q(u)
    =
    \frac{u^q}
         {u^q+(1-u)^q},
    \qquad q>1,
    \label{eq:kress_local_grading}
\end{equation}
where $u$ is given in \cite{kress} and the mapped periodic parameter is
\begin{equation}
    t(\sigma)
    =
    c_j+(c_{j+1}-c_j)g_q(u).
    \label{eq:kress_global_grading}
\end{equation}
The first and second derivatives of this map are evaluated together with the
nodes, and the Jacobian $dt/d\sigma$ is incorporated into the quadrature
weights. Since $g_q'(u)$ vanishes at the endpoints, the discretization is
concentrated near each corner. The grading strength and boundary resolution cannot be varied independently. For a fixed number of points, sufficiently strong grading can produce nearly coincident mapped nodes. The implementation therefore monitors the minimum periodic spacing and progressively reduces the effective grading order if it falls below \texttt{min\_t\_spacing} (otherwise LAPACK segfaults).

For multiply connected billiards, the connected boundary components are
discretized independently. Kress quadrature scheme occurs only for
source--target interactions on the same component, while interactions
between distinct components are handled by the standard quadrature. This also allows the composite boundary-integral formulation to use different discretization strategies on different physical components.

\subsection*{Discrete symmetries}

The symmetry API separates the geometric action from the representation
chosen for a particular calculation. The implemented elementary reflection
types are

\begin{lstlisting}[language=Julia]
XAxisReflection()
YAxisReflection()
XYAxisReflection()
DiagonalReflection()
AntiDiagonalReflection()
\end{lstlisting}
together with arbitrary $n$-fold rotations,
\begin{lstlisting}[language=Julia]
NFoldRotation(n, m)
\end{lstlisting}
representing rotation by $2\pi m/n$. Several reflection generators can be
combined using
\begin{lstlisting}[language=Julia]
CompositeReflection(reflections...)
\end{lstlisting}
and the convenience constructors
\begin{lstlisting}[language=Julia]
D2_symmetry()
Cn_symmetry(n)
\end{lstlisting}
register the complete $D_2$ reflection group and $C_n$ rotational group,
respectively. Each registered transformation receives a stable geometric
identifier. These objects contain only the geometric group action; parity or
irreducible-representation information is supplied separately by the solver.
For boundary-integral methods, symmetry reduction is performed directly on
the discrete periodic boundary. Importantly, transformed coordinates are not
compared numerically to determine corresponding boundary points. Instead,
the registered group algebra and the known reconstruction sectors determine
an exact integer permutation of the midpoint-grid indices. This makes the
discrete symmetry reduction independent of floating-point coordinate
matching.

The node count is chosen to be compatible with the active group so that each
orbit is represented exactly by the periodic midpoint grid. For multiply
connected geometries the orbit construction is applied independently to each
invariant physical boundary component. The current implementation requires
each component to be individually invariant under the active symmetry;
symmetries that exchange distinct physical components are not presently
supported.

\section{Computational complexity}
\label{sec:summary_accelerated}

\begingroup
\small
\renewcommand{\arraystretch}{1.25}
\setlength{\tabcolsep}{6pt}

\begin{xltabular}{\textwidth}{
    >{\raggedright\arraybackslash}p{2.1cm}
    >{\raggedright\arraybackslash}X
    >{\raggedright\arraybackslash}X
}
\caption{
Computational costs per spectral window for the accelerated
eigensolvers in \texttt{QuantumBilliards.jl}.
Window widths are representative choices.
}
\label{tab:eigensolvers}\\

\toprule
\textbf{Solver / window} &
\textbf{Setup} &
\textbf{Subsequent work} \\
\midrule
\endfirsthead

\multicolumn{3}{l}{\tablename~\thetable\ continued}\\
\toprule
\textbf{Solver / window} &
\textbf{Setup} &
\textbf{Subsequent work} \\
\midrule
\endhead

\midrule
\multicolumn{3}{r}{\footnotesize Continued on next page}\\
\endfoot

\bottomrule
\endlastfoot

\textbf{Expanded BIM}\newline
$\Delta k=0.01$--$0.05$
&
$O(N^3+pN^2)$\newline
Matrix and derivative construction;
one LU factorization.
&
$O(tN^2+Nth+c_{\mathrm E}h^3)$\newline
Krylov iterations, orthogonalization,
projected eigensolves, and second-order corrections.
\\

\addlinespace[5pt]

\textbf{Vergini--Saraceno}\newline
$\Delta k=0.01$--$0.05$
&
$O(NM^2)$\newline
Basis evaluation and assembly
of the quadratic-form matrices.
&
$O(M^3)$\newline
Nullspace reduction and the
reduced generalized eigensolve.
\\

\addlinespace[5pt]

\textbf{Beyn}\newline
$\Delta k=0.5$--$1$
&
$O(n_qN^3+n_qpN^2)$\newline
Contour-matrix construction
and $n_q$ LU factorizations.
&
$O(a n_qN^2s+a Ns^2+s^3)$\newline
Contour solves, moment SVDs,
projection, and the reduced eigensolve.
\\

\addlinespace[5pt]

\textbf{CORK}\newline
$\Delta k=0.5$--$1$
&
$O(N^3+p^2N^2)$\newline
Taylor-to-Chebyshev conversion,
initial coefficient-action caching,
and one LU factorization.
&
$\begin{aligned}
O\bigl(&N^2m+pN^2r+pNrm\\
      &{}+prm^2+c_{\mathrm C}m^3\bigr)
\end{aligned}$\newline
Compact Krylov iteration, coefficient-action caching,
orthogonalization, and convergence checks.
\\

\end{xltabular}

{\footnotesize
\noindent
\textit{Parameters.}
$N$: boundary-matrix dimension;
$M$: Helmholtz-basis dimension;
$p$: polynomial degree;
$n_q$: contour-node count.
For expanded BIM, $t$ counts operator applications,
$h$ is the maximum active Krylov dimension,
and $c_{\mathrm E}$ counts projected eigensolves.
For Beyn, $s$ is the maximum probing width
and $a$ counts probing size increase attempts (typically $a=1$; additional attempts are required only when the initial probing size is insufficient).
For CORK, $m$ is the final Krylov dimension,
$r$ is the maximum physical-basis rank,
and $c_{\mathrm C}$ counts convergence checks (at least $2$ are necessary in the current implementation).
\par\smallskip

\noindent
\textit{Assumptions.}
Dense matrices, fixed block size and symmetry multiplicity,
and the analytic Taylor backend.
Krylov costs are upper-bound summaries.
Optional eigenvector reconstruction, additional nonlinear
refinement, and eigenstate postprocessing are excluded.
\par
}

\endgroup

\section{Vergini--Saraceno: basis truncation due to the numerical nullspace}
\label{sec:vs_details}

\begin{figure}[h!]
    \centering
    \includegraphics[width=\linewidth]{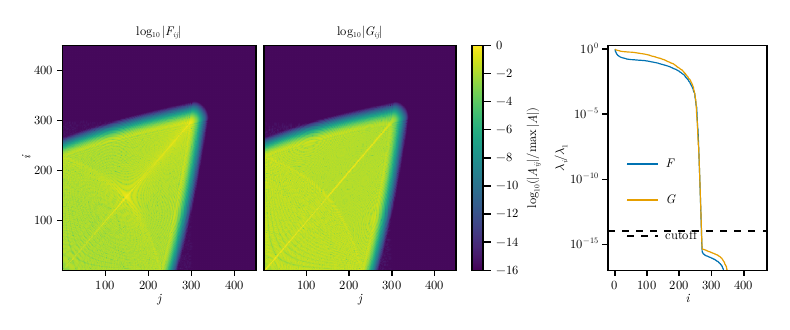}
    \caption{Numerical rank structure of the matrices used in the Decomposition method \eqref{eq:dm_tension} (\texttt{StadiumBilliard(1.0)} geometry with basis \texttt{CornerAdaptedFourierBessel(10, Float64(pi/2),SVector(0.0,0.0),0.0)} and \texttt{k=300.149}. Solver parameters are \texttt{DecompositionMethodSolver(3.0, 10.0)}). The matrix representations of $F$ \eqref{eq:F_matrix} and $G$ (labeled as $N$ in \eqref{eq:N_matrix}) are normalized by their respective maximum absolute values. Owing to the deliberately overcomplete Helmholtz basis, their spectra develop a
    pronounced separation between well-resolved directions and a numerical
    nullspace. The normalized eigenvalue spectra of these matrices show the corresponding sharp
    drop in magnitude. The dashed line indicates the relative numerical cutoff
    used to determine the resolved subspace; directions below this threshold
    are discarded before solving the generalized eigenvalue problem.}
    \label{fig:dm_decomposition}
\end{figure}

The Helmholtz basis used by the Vergini--Saraceno (actually all basis type methods) method is deliberately overcomplete when restricted to the boundary. Consequently, the matrix $F$ in Eq.~\eqref{eq:vs_F_Fk} becomes numerically rank deficient as the
basis size is increased. Retaining the corresponding near-null directions
in Eq.~\eqref{eq:vs_gevp} makes the generalized eigenvalue problem
ill-conditioned and produces unstable results.

The implementation therefore diagonalizes the Hermitian matrix $F$,
\begin{equation}
    F=U\Lambda U^\dagger,
    \label{eq:vs_nullspace_eig}
\end{equation}
and determines its numerical rank from the relative eigenvalue spectrum \cite{VerSar1995,barnett_dissipation}. The resulting separation between the resolved subspace and the numerical
nullspace is illustrated in Fig.~\ref{fig:dm_decomposition}.
If $U_r$ contains the eigenvectors associated with the retained
eigenvalues, the generalized eigenproblem is restricted to this resolved
subspace,
\begin{equation}
    U_r^\dagger F_kU_r\,\mathbf y
    =
    \mu\,
    U_r^\dagger FU_r\,\mathbf y,
    \label{eq:vs_projected_gevp}
\end{equation}
with $\mathbf c=U_r\mathbf y$. The resulting generalized eigenvalues are
then converted to physical wavenumbers using Eq.~\eqref{eq:vs_result}.

The cutoff should lie in the spectral gap separating the resolved subspace
from the numerical nullspace. Its precise value is therefore less important
than the existence of a clear rank separation. In practice, convergence is
verified by increasing the basis and boundary resolutions and checking the
stability of the resulting eigenvalues.

\section{EBIM: accuracy vs. window size}
\label{sec:ebim_details}

\begin{figure}[h!]
    \centering
    \includegraphics[width=0.6\linewidth]{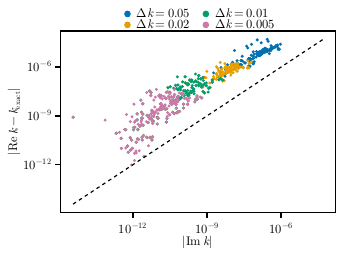}
    \caption{Accuracy of EBIM for a rectangular billiard with side lengths
    $2a=2$ and $2b=2\sqrt[4]{2}$ over the spectral interval
    $95\leq k\leq105$. The spectrum is computed using the second-order
    expansion with boundary resolution \texttt{ppw}=15 and spectral window
    widths $\Delta k=0.05,0.02,0.01,0.005$. The exact analytical
    spectrum allows the error
    $|\operatorname{Re}k-k_{\mathrm{exact}}|$ to be compared directly with
    $|\operatorname{Im}k|$. Decreasing $\Delta k$ systematically reduces
    both quantities, demonstrating the improvement of the local expansion as
    the spectral windows are narrowed. Their correlation further shows that
    $|\operatorname{Im}k|$ provides a useful a posteriori indicator of the
    eigenvalue accuracy (albeit worse than Beyn which does not have visible deterioration of eigenvalue accuracy near the edges). The dashed line indicates
    $|\operatorname{Re}k-k_{\mathrm{exact}}|=|\operatorname{Im}k|$.}
    \label{fig:ebim_rectangle_accuracy}
\end{figure}

The EBIM expansion is local about $k_0$, so only eigenvalues within a
prescribed spectral window are retained. Increasing the window reduces the
number of expansion centers required to cover a spectrum, but increases the
error from the neglected higher-order terms in Eq.~\eqref{eq:ebim_taylor}.
This behavior, and the improvement provided by the second-order correction,
is illustrated in Fig.~\ref{fig:ebim_rectangle_accuracy}.

\section{Beyn: quadrature dependence on numerical rank}
\label{sec:beyn_details}

The accuracy of Beyn's method is controlled by the contour discretization,
the dimension of the probing subspace, and the numerical rank retained in
the moment matrix $A_0$. For the circular contours in
~\eqref{eq:beyn_moments}(later an option for rectangular ones will be added for dielectric cavities \cite{beyn_rectangle_contour}), the contour moments are evaluated using the
periodic trapezoidal rule at

\begin{equation}
    z_j=k_0+Re^{i\theta_j},
    \qquad
    \theta_j=\frac{2\pi j}{N_\Gamma},
    \qquad
    j=0,\ldots,N_\Gamma-1.
\end{equation}

Since the integrands are analytic along the contour provided no singularity
lies on $\Gamma$, the trapezoidal approximation converges exponentially with
$N_\Gamma$ \cite{beyn_contour}. In practice, convergence is checked by increasing the number of contour nodes; approximately $N_\Gamma=40$ is generally sufficient for the calculations considered here.

\begin{figure}[h!]
    \centering
    \includegraphics[width=0.6\linewidth]{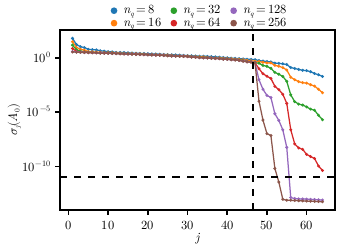}
    \caption{Convergence of the singular values of the Beyn moment matrix
    $A_0$ with the number of contour quadrature nodes $n_q$. The calculation
    is performed for a rectangular billiard with side lengths $2a=2$ and
    $2b=2.914$, using a circular contour centered at $k_0=50$ with radius
    $R=0.5$, enclosing the spectral interval $49.5\leq k\leq50.5$.
    The vertical dashed line marks the exact number of eigenvalues enclosed by
    the contour, while the horizontal dashed line indicates the numerical SVD
    cutoff $\texttt{svd\_tol}=10^{-11}$. In the exact contour integral,
    the rank of \(A_0\) is determined by the enclosed eigenspace. With finite
    quadrature, the nominally vanishing singular values acquire nonzero numerical
    values. As \(n_q\) is increased, these singular values decrease rapidly,
    producing an increasingly sharp separation at the exact rank. This illustrates
    the convergence of the contour quadrature and the resulting stabilization of
    the numerical rank used by Beyn's method.}
    \label{fig:beyn_nq_convergence}
\end{figure}

The contour radius and probing dimension must be chosen together. For a
requested real spectral interval, the implementation uses the Weyl estimate
of the local density of states to construct consecutive contours containing
a prescribed number of expected eigenvalues, subject to the maximum radius
\texttt{Rmax}. Keeping the contours sufficiently small is also important for
maintaining a clear separation between the significant and numerical
singular values of $A_0$.

The number of eigenvalues actually enclosed by a contour is determined from
the singular values
\begin{equation}
    \sigma_1\geq\sigma_2\geq\cdots
\end{equation}
of $A_0$. The numerical rank is obtained by retaining singular values above
the relative cutoff
\begin{equation}
    \frac{\sigma_j}{\sigma_1}>\texttt{svd\_tol}.
    \label{eq:beyn_rank}
\end{equation}
The spectrum of singular-values of $A_0$ is illustrated in
Fig.~\ref{fig:beyn_nq_convergence}. The probing dimension must exceed this numerical rank. If the retained rank approaches the available probing dimension, the latter is increased and the
contour calculation is repeated. This adaptive procedure avoids choosing an
unnecessarily large probing matrix for every contour while reducing the risk
of missing eigenvalues in locally dense parts of the spectrum.

\begin{figure}[h!]
    \centering
    \includegraphics[width=0.6\linewidth]{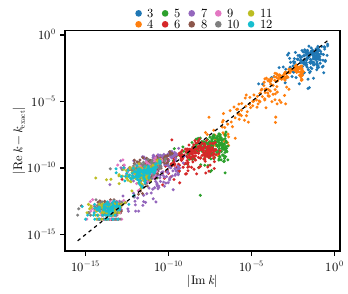}
    \caption{Relation between the imaginary part of the wavenumber and the
    actual eigenvalue error for the rectangular test problem over
    $90\leq k\leq100$, computed with Beyn's method (\texttt{DoubleLayerPotentialSolver(ppw, grading = GlobalCornerGrading(kressq=2), sym\_characters = (-1, -1)}) at increasing boundary
    resolutions \texttt{ppw} (the value in the legend). Due to exponential convergence of the trapezoidal rule. \texttt{nq=40} was enough to rule out the deterioration of eigenvalue accuracy due to poorly resolved contour quadrature. The exact analytical spectrum allows the error
    $|\operatorname{Re}k-k_{\mathrm{exact}}|$ to be compared directly with
    $|\operatorname{Im}k|$. Increasing the number of points per wavelength
    (\texttt{ppw}) decreases both quantities, while their close correlation over
    many orders of magnitude shows that $|\operatorname{Im}k|$ provides a
    useful a posteriori estimate of the eigenvalue error when the exact spectrum
    is unavailable. The dashed line indicates
    $|\operatorname{Re}k-k_{\mathrm{exact}}|=|\operatorname{Im}k|$.
    The two visible bands at a given resolution arise from the arithmetic
    degeneracies of the rectangle: Beyn resolves the degenerate eigenvalues
    accurately, although these pairs typically acquire somewhat larger imaginary
    parts than non-degenerate levels.}
    \label{fig:beyn_accuracy}
\end{figure}

After solving the reduced eigenproblem in Eq.~\eqref{eq:beyn_reduced},
candidate eigenpairs are checked directly against the original nonlinear
Fredholm problem. For a candidate $k_\alpha$ and reconstructed boundary
vector $\mathbf v_\alpha$, the residual
\begin{equation}
    \rho_\alpha
    =
    \frac{\|A(k_\alpha)\mathbf v_\alpha\|_2}
         {\|\mathbf v_\alpha\|_2}
    \label{eq:beyn_residual}
\end{equation}
is evaluated and compared with \texttt{res\_tol}. This provides an
independent a posteriori test of the contour reduction and removes candidates
associated with an ambiguous numerical rank or insufficient contour
resolution. In production runs heuristics are used that circumvent the calculations of all residuals by just checking those with the largest imaginary part. The imaginary part of the computed wavenumber also provides a useful a posteriori diagnostic of the eigenvalue accuracy, as illustrated in Fig.~\ref{fig:beyn_accuracy}. The final accuracy is ultimately limited by the discretization of the underlying boundary-integral operator.

Two possible extensions are being investigated to reduce the computational
cost of the contour solves. The first is a hierarchically semi-separable
(HSS) representation of the boundary-integral matrices~\cite{hss} (or a modern rank structured format 
that can handle strongly oscillating kernels at higher $k$).
The dense Fredholm matrices contain compressible off-diagonal interactions,
so an HSS backend could reduce both the memory requirement and the cost of
the repeated factorizations required at the contour nodes.

\subsection{Half-contour approach}
\label{sec:half_contour}

A second possibility is a half-contour Beyn search using a real
Green function with an added homogeneous
solution~\cite{backer_BIM,Heller_GFM},
\begin{equation}
    G_k^{(\beta)}(\mathbf x,\mathbf y)
    =
    -\frac14\left[
        Y_0(k|\mathbf x-\mathbf y|)
        +\beta J_0(k|\mathbf x-\mathbf y|)
    \right],
    \qquad \beta\in\mathbb R,
    \label{eq:beyn_real_green_beta}
\end{equation}
with fixed real $\beta$. This formulation is known to introduce auxiliary
roots \cite{backer_BIM}, but Beyn can circumvent this.
Let $\Gamma=\partial\Omega$ be a smooth curve with the unit normal denoted by $\mathbf{n}$.
For fixed distinct points $\mathbf{s,x}\in\Omega$ in the interior the Green functions satisfies
$(\Delta_{\mathbf{y}}+k^2)G_k^{(\beta)}(\mathbf{s},\mathbf{y})=-\delta(\mathbf{s}-\mathbf{y})$ and
$(\Delta_{\mathbf{y}}+k^2)G_k^{(D)}(\mathbf{x},\mathbf{y})=-\delta(\mathbf{x}-\mathbf{y})$, with the billiard's Green function $G_k^{(D)}(\mathbf{x},\mathbf{z})=0$ for $\mathbf{z}\in\Gamma$. By Green's identity \cite{kress_book} we have
\begin{align}
I
&=\iint_\Omega
\left[
G_k^{(\beta)}(\mathbf{s},\mathbf{y})(\Delta_{\mathbf{y}}+k^2)G_k^{(D)}(\mathbf{x},\mathbf{y})
-G_k^{(D)}(\mathbf{x},\mathbf{y})(\Delta_{\mathbf{y}}+k^2)G_k^{(\beta)}(\mathbf{s},\mathbf{y})
\right]\,d\mathbf{y}
\nonumber\\
&=-G_k^{(\beta)}(\mathbf{s},\mathbf{x})+G_k^{(D)}(\mathbf{x},\mathbf{s})
=G_k^{(D)}(\mathbf{s},\mathbf{x})-G_k^{(\beta)}(\mathbf{s},\mathbf{x}),
\label{eq:green_th_volume}
\end{align}
where we used the Green function symmetry $G(\mathbf{x},\mathbf{y})=G(\mathbf{y},\mathbf{x})$. As always this is then connected to its contour integral connecting formula 
\begin{align}
I
&=\oint_\Gamma
\left[
G_k^{(\beta)}(\mathbf{s},\mathbf{z})\partial_{\mathbf{n_z}}G_k^{(D)}(\mathbf{x},\mathbf{z})
-G_k^{(D)}(\mathbf{x},\mathbf{z})\partial_{\mathbf{n_z}}G_k^{(\beta)}(\mathbf{s},\mathbf{z})
\right]\,d\ell_\mathbf{z}
\nonumber\\
&=\oint_\Gamma G_k^{(\beta)}(\mathbf{s},\mathbf{z})u_{\mathbf{x}}(\mathbf{z};k)\,d\ell_{\mathbf{z}},
\qquad
u_{\mathbf{x}}(\mathbf{z},k):=\partial_{\mathbf{n_z}}G_k^{(D)}(\mathbf{x},\mathbf{z}).
\label{eq:green_th_boundary}
\end{align}
where the second term drops due to $G_k^{(D)}(\cdot,\mathbf{z}\in\Gamma)=0$. Consequently, for $\mathbf{s,x}\in\Omega$,
\begin{equation}
G_k^{(D)}(\mathbf{s},\mathbf{x})
=
G_k^{(\beta)}(\mathbf{s},\mathbf{x})
+\oint_\Gamma G_k^{(\beta)}(\mathbf{s},\mathbf{z})u_{\mathbf{x}}(\mathbf{z},k)\,d\ell_{\mathbf{z}}.
\label{eq:green_D_representation}
\end{equation}
Taking the interior trace as $\mathbf{s}$ approaches a point on the boundary $\mathbf{s}\to \mathbf{z}\in\Gamma$ and using the jump relation on the used interior normal derivative at that point on \eqref{eq:green_D_representation} we get
\begin{align}
u_\mathbf{x}(\mathbf{z},k)
&=\partial_{\mathbf{n_z}}G_k^{(D)}(\mathbf{z},\mathbf{x})
\nonumber\\
&=\lim_{\varepsilon\downarrow0}
\left.
\mathbf n(\mathbf{z})\cdot\nabla_\mathbf{s}
\left[
G_k^{(\beta)}(\mathbf{s},\mathbf{x})
+\oint_\Gamma G_k^{(\beta)}(\mathbf{s},\mathbf{v})u_\mathbf{x}(\mathbf{v},k)\,d\ell_\mathbf{v}
\right]
\right|_{\mathbf{s}=\mathbf{z}-\varepsilon\mathbf n(\mathbf{z})}
\nonumber\\
&=\partial_{\mathbf{n_z}}G_k^{(\beta)}(\mathbf{z},\mathbf{x})
+\frac12u_\mathbf{x}(\mathbf{z},k)
+\oint_\Gamma
\partial_{\mathbf{n_z}}G_k^{(\beta)}(\mathbf{z},\mathbf{v})u_\mathbf{x}(\mathbf{v},k)\,d\ell_\mathbf{v},
\end{align}
where the last equality uses the jump relation for the normal derivative of the single-layer potential. Collecting like terms gives
\begin{align}
[A_\beta(k)u_\mathbf{x}](\mathbf{z})
&=u_\mathbf{x}(\mathbf{z},k)
-2\oint_\Gamma
\partial_{\mathbf{n_z}}G_k^{(\beta)}(\mathbf{z},\mathbf{v})u_\mathbf{x}(\mathbf{v};k)\,d\ell_\mathbf{v}=2\partial_{\mathbf{n_z}}G_k^{(\beta)}(\mathbf{z},\mathbf{x}).
\label{eq:green_probe_equation}
\end{align}
where $A_\beta(k)=I-2K'_\beta(k)$ will lead to the discretized (jump-scaled) Fredholm matrix $A_{\beta,ij}$. This result is the key to removing spurious solutions, as choosing for the probing matrix $V_{\beta,ij}$
\begin{equation}
V_{\beta,ij}(k)
=
2\partial_{\mathbf{n_{z_i}}}G_k^{(\beta)}(\mathbf{z_i},\mathbf{x_j}).
\end{equation}
will lead to the zeroth order matrix element (easily shown by inserting \eqref{eq:green_probe_equation})
\begin{equation}
    [A_\beta(k)^{-1}V_\beta(k)]_{ij}
    =
    \sum_{\ell=1}^{N}
    [A_\beta(k)^{-1}]_{i\ell}
    \,2\partial_{\mathbf{n_{z_\ell}}}G_k^{(\beta)}(\mathbf{z_\ell},\mathbf{x_j})
    =
    \partial_{\mathbf{n_{z_i}}}G_k^{(D)}(\mathbf{z_i},\mathbf{x_j}).
    \label{eq:green_probe_response}
\end{equation}
Thus the zeroth order Beyn moment over the contour $C$ is (after inserting \eqref{eq:green_probe_response})
\begin{equation}
    (M_0)_{ij}
    =
    \frac{1}{2\pi i}\oint_C
    [A_\beta(k)^{-1}V_\beta(k)]_{ij}\,dk
    =
    \frac{1}{2\pi i}\sum_m\oint_C
    \frac{\partial_n\psi_m(\mathbf{z_i})\,\overline{\psi_m(\mathbf{x_j})}}{k_m^2-k^2}\,dk.
\end{equation}
Here $\psi_m$ are orthonormal Dirichlet eigenfunctions, coming from the "sum over eigenstates" definition of $G_D$. This has only physical Dirichlet poles (provided we sample well enough s.t. not all $\psi_m(\mathbf{x_j})$ are nodal lines - giving no moment accumulation). Using this approach the auxiliary singularities of $A_\beta(k)^{-1}$ are removable. Both $A_\beta$ and $V_\beta$ must however obey complex-conjugation symmetry (therefore native CFIE approach is not directly applicable)
\begin{equation}
    X_\beta(\overline{k})=\overline{X_\beta(k)},
    \qquad X_\beta=A_\beta^{-1}V_\beta.
    \label{eq:beyn_conjugation}
\end{equation}
Therefore a conjugate-paired \texttt{nq}-node Beyn rule
requires only $\texttt{nq}/2$ independent contour factorizations. This experimental feature is accesed using the \texttt{use\_half\_contour=true} kwarg.

\section{Analytic construction of the Taylor polynomial of $A(k)$}
\label{sec:cork_polynomial}

The CORK solver (and all BIM solvers with \texttt{use\_taylor=true}) backend requires a polynomial approximation of the Fredholm matrix
$A(k)$ over a spectral window of size $\Delta$. Rather than evaluating the complete
boundary-integral matrix at a set of interpolation points, the implementation
constructs this polynomial from an analytic Taylor expansion about the window
center $A(k_0)$. The dependence of the DLP and CFIE kernels using the Kress quadrature scheme on $k$ is carried by a small number of additional Bessel and Hankel functions, whose Taylor coefficients can be generated efficiently by recurrence.

For a given interval
\[
    k\in[k_0-\Delta,k_0+\Delta],
\]
the polynomial is constructed on a slightly larger interval with half-width
\[
    \Delta_p=(1+g)\Delta,
\]
where $g$ is the guard parameter as described in Sec.\ref{sec:cork}. Introducing
\[
    \delta=k-k_0,
    \qquad
    t=\frac{\delta}{\Delta_p},
\]
we first write
\begin{equation}
    A(k_0+\delta)
    =
    \sum_{\ell=0}^{p}A_\ell\delta^\ell
    +O(\delta^{p+1}),
    \qquad
    A_\ell=\frac{A^{(\ell)}(k_0)}{\ell!}.
    \label{eq:cork_taylor}
\end{equation}
For the DLP kernel, the off-diagonal Kress quadrature scheme splitting contains the
radial functions $kJ_1(kr)$ and $kH_1^{(1)}(kr)$. Considering first
\[
    f(k)=kJ_1(kr).
\]
The Bessel equation implies \cite{gradshteyn_ryzhik}
\begin{equation}
    f''(k)-\frac{1}{k}f'(k)+r^2f(k)=0.
    \label{eq:cork_j1_ode}
\end{equation}
Writing
\[
    f(k_0+\delta)=\sum_{n=0}^{\infty}f_n\delta^n,
    \qquad
    f_n=\frac{f^{(n)}(k_0)}{n!},
\]
and matching equal powers of $\delta$ gives (including $k=k_0+\delta$)
\begin{equation}
    f_{n+2}
    =
    -\frac{
        (n+1)(n-1)f_{n+1}
        +k_0r^2f_n+r^2f_{n-1}}
        {k_0(n+2)(n+1)},
    \qquad n\geq0,
    \label{eq:cork_j1_rec}
\end{equation}
with
\begin{equation}
    f_{-1}\equiv0,
    \qquad
    f_0=k_0J_1(k_0r),
    \qquad
    f_1=k_0rJ_0(k_0r).
    \label{eq:cork_j1_initial}
\end{equation}
The same recurrence applies to

\[
    h(k)=kH_1^{(1)}(kr),
\]

with
\begin{equation}
    h_{-1}\equiv0,
    \qquad
    h_0=k_0H_1^{(1)}(k_0r),
    \qquad
    h_1=k_0rH_0^{(1)}(k_0r).
    \label{eq:cork_h1_initial}
\end{equation}
Thus only the initial $H_0^{(1)}(k_0r),H_1^{(1)}(k_0r)$ and $J_1(k_0r)$ functions are evaluated explicitly; all higher Taylor coefficients are obtained by scalar recurrence.

To see how these coefficients enter the discretized operator, recall the
Kress splitting of the parametrized DLP kernel \cite{kress},
\begin{equation}
    L(t,\tau)
    =
    L_1(t,\tau)
    \log\!\left(
        4\sin^2\frac{t-\tau}{2}
    \right)
    +L_2(t,\tau).
    \label{eq:cork_kress_L}
\end{equation}
The logarithmic coefficient $L_1$ is proportional to $kJ_1(kr)$, while
\begin{equation}
    L_2(t,\tau)
    =
    L(t,\tau)
    -
    L_1(t,\tau)
    \log\!\left(
        4\sin^2\frac{t-\tau}{2}
    \right)
\end{equation}
contains both the $kH_1^{(1)}(kr)$ contribution of the original kernel and
the extra $kJ_1(kr)$ term \cite{kress}. Consequently, after Nystr\"om
discretization an off-diagonal matrix entry has the form
\begin{equation}
    D_{ij}(k)
    =
    R_{ij}L_{1,ij}(k)
    +w_jL_{2,ij}(k),
    \qquad i\neq j,
    \label{eq:cork_kress_dlp}
\end{equation}
where $R_{ij}$ denotes the Kress quadrature weight and $w_j$
the ordinary trapezoidal quadrature weight. Since all remaining factors depend only on
the boundary geometry, the Taylor coefficients of
$D_{ij}(k)$ follow directly from the recurrences of $kJ_1(kr)$ and
$kH_1^{(1)}(kr)$.

On the diagonal, the Kress limit of the double-layer kernel is independent
of $k$,
\begin{equation}
    D_{ii}=w_i\kappa_i,
    \label{eq:cork_dlp_diag}
\end{equation}
and therefore the diagonal contributes only to the zeroth Taylor
coefficient.

For the CFIE,
\begin{equation}
    A(k)=I-\left[D(k)+ikS(k)\right],
    \label{eq:cork_cfie}
\end{equation}
the SLP contribution introduces the additional radial functions
$kJ_0(kr)$ and $kH_0^{(1)}(kr)$ (the factor $k$ multiplying the
off-diagonal SLP kernel originates from the choice $\eta(k)=k$ in the
CFIE \cite{kress,fredholm_determinant_method}). Tipically it is written as
\begin{equation}
    M(t,\tau)
    =
    M_1(t,\tau)
    \log\!\left(
        4\sin^2\frac{t-\tau}{2}
    \right)
    +M_2(t,\tau),
    \label{eq:cork_kress_M}
\end{equation}
where $M$ is proportional to $H_0^{(1)}(kr)$, $M_1$ is proportional
to $J_0(kr)$, and $M_2$ contains both the Hankel contribution and the
compensating Bessel term. Multiplication by the $ik$ \eqref{eq:cork_cfie}
therefore leaves the off-diagonal SLP contribution
expressible in terms of $kJ_0(kr)$ and $kH_0^{(1)}(kr)$.

For $y(k)=J_0(kr)$ \cite{gradshteyn_ryzhik},
\begin{equation}
    y''(k)+\frac{1}{k}y'(k)+r^2y(k)=0,
    \qquad
    y(k_0+\delta)=\sum_{n=0}^{\infty}y_n\delta^n.
\end{equation}
The Taylor coefficients satisfy (including $k=k_0+\delta$)
\begin{equation}
    y_{n+2}
    =
    -\frac{
        (n+1)^2y_{n+1}
        +k_0r^2y_n+r^2y_{n-1}}
        {k_0(n+2)(n+1)},
    \qquad n\geq0,
    \label{eq:cork_j0_rec}
\end{equation}
with
\begin{equation}
    y_{-1}\equiv0,
    \qquad
    y_0=J_0(k_0r),
    \qquad
    y_1=-rJ_1(k_0r).
\end{equation}
If $q_n$ denotes the Taylor coefficient of $kJ_0(kr)$, multiplication by
$k=k_0+\delta$ gives
\begin{equation}
    q_0=k_0y_0,
    \qquad
    q_n=k_0y_n+y_{n-1},
    \quad n\geq1.
    \label{eq:cork_kj0}
\end{equation}
The same applies to $H_0^{(1)}(kr)$. In this case,
\begin{equation}
    y_{-1}\equiv0,
    \qquad
    y_0=H_0^{(1)}(k_0r),
    \qquad
    y_1=-rH_1^{(1)}(k_0r),
\end{equation}
and \eqref{eq:cork_j0_rec} generates the higher Taylor coefficients.
The coefficients of $kH_0^{(1)}(kr)$ are obtained from
\eqref{eq:cork_kj0} by the same product rule.

The diagonal of CFIE's SLP contribution requires separate
treatment, as the finite diagonal limit of $M_1$ retains an
explicit logarithmic dependence on $k$. To make the geometric factors
explicit, define $s(t)$ as the norm of the tangent at parameter $t$
\begin{equation}
    s(t)=|\mathbf{x}'(t)|
    =
    \sqrt{[x_1'(t)]^2+[x_2'(t)]^2},
    \qquad ds=s(t)\,dt
    \label{eq:cork_metric_factor}
\end{equation}
The coefficient multiplying the log of the SLP
kernel is
\begin{equation}
    M_1(t,\tau)
    =
    -\frac{1}{2\pi}
    J_0\!\left(kr(t,\tau)\right)s(\tau).
\end{equation}
Since $r(t,t)=0$ and $J_0(0)=1$, its diagonal value at the quadrature node $t_i$, (writing $s_i=s(t_i)$) is
\begin{equation}
    M_1(t,t)=-\frac{s(t)}{2\pi}
    \rightarrow
    -\frac{s_i}{2\pi}= M_{1,ii}.
\end{equation}
Using the diagonal limit as defined in \cite{kress} (and simplifying $-\frac{1}{2\pi}\ln\left[\left(\frac{k s(t)}{2}\right)^2\right]=-\frac{1}{\pi}\ln\left(\frac{k s(t)}{2}\right)$),
\begin{equation}
    M_{2,ii}(k)
    =
    s_i\left[
        \frac{i}{2}
        -\frac{\gamma}{\pi}
        -\frac{1}{\pi}\log\left(\frac{k s_i}{2}\right)
    \right],
    \label{eq:cork_M2_diag}
\end{equation}
where $\gamma$ is Euler's constant. Applying the logarithmic quadrature to
$M_1$ and the trapezoidal quadrature to $M_2$ gives
\begin{equation}
    S_{ii}(k)
    =
    R_{ii}M_{1,ii}
    +w_iM_{2,ii}(k),
\end{equation}
and therefore
\begin{align}
    S_{ii}(k)
    &=
    -\frac{R_{ii}s_i}{2\pi}
    +w_i s_i
    \left[
        \frac{i}{2}
        -\frac{\gamma}{\pi}
        -\frac{1}{\pi}\log\left(\frac{k s_i}{2}\right)
    \right] \nonumber\\
    &=
    -\frac{R_{ii}s_i}{2\pi}
    +w_i s_i
    \left[
        \frac{i}{2}
        -\frac{\gamma}{\pi}
        +\frac{1}{2\pi}\log\frac{4}{s_i^2}
        -\frac{1}{\pi}\log k
    \right].
    \label{eq:cork_S_diag_expanded}
\end{align}
\noindent The explicit form of $R_{ii}$ is given by (Eq. (3.4) in \cite{kress} with $k=0$, for an even number of boundary collocation points $N=2n$)
\begin{equation}
    [R_{k=0}^{N=2n}]_{ii}
    \equiv R_{ii} = 
    -\frac{2\pi}{n}
    \sum_{m=1}^{n-1}\frac1m
    -\frac{\pi}{n^2}
    =
    -\frac{2\pi}{n}H_{n-1}
    -\frac{\pi}{n^2},
    \label{eq:kress_R_ii}
\end{equation}
where $H_n$ is the $n$-th Harmonic number. Separating the $k$-dependent term, this can be written compactly as
\begin{equation}
    S_{ii}(k)=c_i-a_i\log k,
    \label{eq:cork_s_diag}
\end{equation}
where
\begin{equation}
    a_i=\frac{w_i s_i}{\pi},
    \qquad
    c_i
    =
    s_i\left[
        -\frac{R_{ii}}{2\pi}
        +w_i\left(
            \frac{i}{2}
            -\frac{\gamma}{\pi}
            +\frac{1}{2\pi}\log\frac{4}{s_i^2}
        \right)
    \right].
    \label{eq:cork_ai_ci}
\end{equation}
\begin{figure}[h!]
    \centering
    \includegraphics[width=0.6\linewidth]{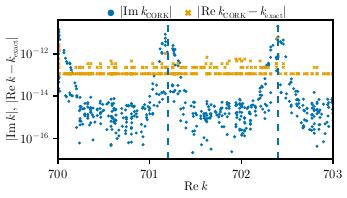}
    \caption{Accuracy of eigenvalues obtained with CORK for a rectangular
    billiard with side lengths \(2a=2\) and \(2b=2\sqrt{2}\), using
    \texttt{GlobalCornerGrading(kressq=2)} as defined in
    Appendix~\ref{sec:geometry}, over
    \(700\leq\operatorname{Re}k\leq703\). The real-part error is determined
    from the exact analytical spectrum, while \(|\operatorname{Im}k|\)
    provides an independent proxy accuracy diagnostic. Vertical dashed lines indicate the individual approximation
    intervals. For \(p=14\), \(|\operatorname{Im}k|\) is the smallest near the centers of the windows and increases towards their edges, where the local polynomial approximation is the least accurate. The real-part error, however, reaches a floor of approximately \(k\epsilon_{\rm mach}\simeq1.6\times10^{-13}\), corresponding to the floating-point resolution at \(k\simeq700\). Thus, away from the window edges, the CORK approximation is sufficiently accurate that the error in \(\operatorname{Re}k\) is limited by double-precision arithmetic. The attainable accuracy near the window edges is instead controlled by the local Chebyshev approximation.}
    \label{fig:cork_window_accuracy}
\end{figure}

Since we use $\eta(k)=k$ \cite{kress,kress_book} in \eqref{eq:cork_cfie}, the diagonal entry of the combined operator $K(k)=D(k)+ikS(k)$ is
\begin{equation}
    K_{ii}(k)
    =
    w_i\kappa_i
    +ikc_i-ia_i k\log k.
    \label{eq:cork_cfie_diag}
\end{equation}
The first two terms are respectively constant and linear in $k$;
therefore, beyond first order, the Taylor expansion is determined entirely
by $k\log k$. The zeroth and first Taylor coefficients about
$k_0$ are
\begin{align}
    K_{ii,0}
    &=
    w_i\kappa_i
    +ik_0\left(c_i-a_i\log k_0\right),
    \label{eq:cork_cfie_diag0}\\
    K_{ii,1}
    &=
    i\left[c_i-a_i(\log k_0+1)\right].
    \label{eq:cork_cfie_diag1}
\end{align}
For $n\geq2$,
\begin{equation}
    \frac{d^n}{dk^n}(k\log k)
    =
    (-1)^n\frac{(n-2)!}{k^{n-1}},
\end{equation}
and hence, after division by $n!$,
\begin{equation}
    K_{ii,n}
    =
    ia_i\frac{(-1)^{n+1}}
    {n(n-1)k_0^{\,n-1}},
    \qquad n\geq2.
    \label{eq:cork_cfie_diagn}
\end{equation}
Thus the remaining logarithmic $k$-dependence of the CFIE diagonal is
included analytically in the Taylor polynomial.

For multiply connected billiards, inter-component interactions are smooth and
their Taylor coefficients follow directly from the
$kH_1^{(1)}(kr)$ recurrence for the DLP and additionally from
$kH_0^{(1)}(kr)$ for the CFIE.

Once the Taylor coefficients of the boundary operator $K(k)$ have been
assembled, those of the Fredholm matrix $A(k)=I-K(k)$ are simply
\begin{equation}
    A_0=I-K_0,
    \qquad
    A_\ell=-K_\ell,\quad \ell\geq1,
\end{equation}
where $A_\ell$ is the $\ell$'th Taylor coefficient. Substituting $\delta=\Delta_p t$, the Taylor polynomial is converted to the Chebyshev basis using
\begin{equation}
    \Delta_p^\ell t^\ell
    =
    \sum_{j=0}^{\ell}C_{j\ell}T_j(t),
    \qquad
    B_j=\sum_{\ell=0}^{p}C_{j\ell}A_\ell,
\end{equation}
where the coefficients $C_{j\ell}$ are generated recursively from
$tT_0=T_1$ and $tT_j=(T_{j-1}+T_{j+1})/2$. This gives the matrix
polynomial
\begin{equation}
    P(t)=\sum_{j=0}^{p}B_jT_j(t)
    \simeq A(k_0+\Delta_p t).
    \label{eq:cork_final_poly}
\end{equation}
The local accuracy of this approximation is illustrated in
Fig.~\ref{fig:cork_window_accuracy}. The polynomial eigenvalue problem
in \eqref{eq:cork_pep} is solved using the structured Chebyshev linearization
\cite{CORK_practical,CORK}. Introducing
\begin{equation}
    z_j=T_j(t)u,\qquad j=0,\ldots,p-1,
\end{equation}
the three-term recurrence gives
\begin{equation}
    z_1=tz_0,\qquad
    z_{j+1}+z_{j-1}=2tz_j,\qquad j=1,\ldots,p-2.
\end{equation}
Eliminating the final Chebyshev term,
$T_p(t)u=2tz_{p-1}-z_{p-2}$, gives
\begin{equation}
    \sum_{j=0}^{p-3}B_jz_j
    +(B_{p-2}-B_p)z_{p-2}
    +B_{p-1}z_{p-1}
    =
    -2tB_pz_{p-1}.
\end{equation}
Together these equations define the generalized eigenvalue problem
\begin{equation}
    L_0z=tL_1z,
    \label{eq:cork_linearization}
\end{equation}
of formal dimension $pN$. The matrices $L_0$ and $L_1$ are not formed
explicitly \cite{CORK_practical}. Instead, shift--invert iteration at $t=0$ is applied to
$M=L_0^{-1}L_1$. Using $T_{2q}(0)=(-1)^q$ and
$T_{2q+1}(0)=0$, each application of $M$ reduces to the Chebyshev block
recurrence and one solve with
\begin{equation}
    P(0)=\sum_{j=0}^{p}B_jT_j(0)
        =B_0-B_2+B_4-\cdots.
    \label{eq:cork_P0}
\end{equation}
Thus $P(0)$ is factorized only once per spectral window and the
factorization is reused throughout the Krylov iteration.

The $pN$-dimensional Krylov vectors are stored in compact form by expressing
their physical blocks in a common orthonormal basis $U$,
\begin{equation}
    z_j=Ug_j,
    \qquad
    W_j=B_jU,
\end{equation}
where the coefficient actions $W_j$ are cached. Thus a linearized vector
$z=[z_0;\ldots;z_{p-1}]$ is represented by the common physical basis $U$
and the corresponding compact coefficient vectors $g_j$. This compact
representation avoids explicitly storing the full $pN$-dimensional Krylov
basis and substantially reduces the associated storage and orthogonalization
costs~\cite{CORK_practical,CORK}.

The resulting compact block-Krylov iteration is continued until the computed
roots in the requested spectral window are stable and satisfy the prescribed residual
\texttt{res\_tol} and imaginary-part tolerances \texttt{imag\_tol}. The corresponding wavenumbers are
finally recovered using \eqref{eq:cork_recovery}.

\section{Future plans}
\label{sec:appendix_future}

The current section is under thorough testing before release. For a billiard on the hyperbolic plane with Gaussian curvature $-1$
the energy is expressed as $E=k^2+\frac{1}{4}$ and the Green function is
\begin{equation}
    G_H(\mathbf x,\mathbf y;k)
    =\frac{1}{2\pi}Q_{-1/2-ik}
      \bigl(\cosh d_H(\mathbf x,\mathbf y)\bigr),
      \label{eq:green_hyperbolic}
\end{equation}
where $d_H$ is the hyperbolic distance and $Q_a(x)$ the associated Legendre function of the second kind \cite{aurich_equation_wavefunction_hyperbolic}. Its analytic Kress split and radial recurrence
follow Ref.~\cite{Orel2026Hyperbolic}, Appendices 1 and 3.

For the magnetic problem the Hamiltonian $H$ in a homogenous magnetic field $B$ (expressed through the magnetic vector potential $\mathbf{A}$) can be written as
\begin{equation}
H=(-i\nabla-\mathbf A)^2, \qquad
\mathbf A=B(-x_2,x_1)/2
\end{equation}
where energy is $E=k^2$. The classical cyclotron radius is $\rho=k/B$, so $2\rho$ is the cyclotron diameter
and the turning point of the radial equation \cite{Hornberger_Smilansky}. The calculations are performed at fixed
$\rho$, with $B=k/\rho$ increasing with $k$. The Green function is \cite{Hornberger_Smilansky}
\begin{equation}
    G_M(\mathbf x,\mathbf y;k)
    =
    \frac{e^{i\phi_B(k)}}{4\pi}
    \Gamma(\frac12-\nu(k))e^{-z/2}U(\frac12-\nu(k),1,z),
    \qquad
    z=\frac{B(k)}{2}|\mathbf x-\mathbf y|^2,
    \label{eq:green_mag}
\end{equation}
where the parameter $\nu$ and the magnetic phase $\phi_B(k)$ are defined following \cite{Hornberger_Smilansky}
\begin{equation}
    \nu(k)=\frac{k^2}{2B(k)},
    \qquad
    \phi_B(k)=\frac{B(k)}{2}(y_1x_2-y_2x_1).
\end{equation}
Here $U$ is Tricomi's confluent hypergeometric function. Constructing the Fredholm matrix directly via \eqref{eq:green_mag} causes issues with both Beyn's method and CORK due to the presence of Landau poles. They are regularized following \cite{Hornberger_Smilansky} giving $\widetilde G_M=\beta G_M$, with $\beta(\nu)=\cos(\pi\nu)$ and explicitly given by
\begin{align}
    \widetilde G_M
    &=
    \frac{e^{i\phi_B(k)-z/2}}
         {4\Gamma(\frac12+\nu(k))}
    U\!\left(\tfrac12-\nu(k),1,z\right),
    \\
    \widetilde G_M\big|_{\nu=\nu_n}
    &=
    \frac{(-1)^n}{4}
    e^{i\phi_{B_n}-z_n/2}L_n(z_n),
    \qquad \nu_n=n+\tfrac12,
    \label{eq:regularized_green_mag}
\end{align}
where $n=0,1,\ldots$, $L_n$ is the Laguerre polynomial \cite{NIST:DLMF}, $B_n=(2n+1)/\rho^2$, and $z_n=B_n|\mathbf x-\mathbf y|^2/2$. This follows by inserting $\nu_n=n+\tfrac12$ and using the Tricomi–Laguerre identity $U(-n,1,z)=(-1)^n n!L_n(z)$ (Eq.~13.6.19 in \cite{NIST:DLMF}). The zero of $\cos(\pi\nu)$ cancels the free-space Landau pole, giving the finite analytic continuation \eqref{eq:regularized_green_mag}. The corresponding regularized CFIE Fredholm operator condition is
\begin{equation}
    \widetilde{A}_M\sigma
    =
    \left(-\frac{\beta(\nu)}2I
    +\widetilde K'_M+ik\widetilde S_M\right)\sigma
    =
    -\frac{\beta(\nu)}2\sigma(\mathbf x)
    +\int_{\partial\Omega}
    \widetilde h_M(\mathbf x,\mathbf y;k)
    \sigma(\mathbf y)\,ds_{\mathbf y}=0,
    \label{eq:mag_cfie_fredholm_reg}
\end{equation}
where $\widetilde S_M$ and $\widetilde K'_M$ denote the regularized magnetic
single-layer and adjoint double-layer operators and $
\widetilde h_M(\mathbf x,\mathbf y;k)
=(D_{n_{\mathbf x}}+ik)\widetilde G_M(\mathbf x,\mathbf y;k)$ is the combined kernel and 
$ D_{n_{\mathbf x}}
=\partial_{n_{\mathbf x}}
-i\mathbf n_{\mathbf x}\cdot\mathbf A(\mathbf x)
$ is the normal covariant derivative. Using $\partial_z U(\frac12-\nu(k),1,z)=-(\frac12-\nu(k))U(\frac32-\nu(k),2,z)$ from Eq.13.3.22 from \cite{NIST:DLMF} it can be expressed as
\begin{align}
    \widetilde h_M
    ={}&
    \frac{e^{i\phi_B-z/2}}{4\Gamma(\tfrac12+\nu)}
    \Bigg\{
    -B\,\mathbf n_{\mathbf x}\cdot(\mathbf x-\mathbf y)
    \left[(\tfrac12-\nu(k))U(\tfrac32-\nu(k),2,z)+\frac12U(\tfrac12-\nu(k),1,z)\right]
    \nonumber\\
    &\hspace{3em}
    +i\left[k+\mathbf n_{\mathbf x}\cdot
    \bigl(\mathbf A(\mathbf y)-\mathbf A(\mathbf x)\bigr)\right]
    U(\tfrac12-\nu(k),1,z)
    \Bigg\}.
\end{align}
Using $\frac{\partial}{\partial z}L_n(z)=-L_{n-1}^{(1)}(z)$
(Eq.~18.9.23 in \cite{NIST:DLMF}), its Landau limit is
\begin{align}
    \widetilde h_M\big|_{\nu_n}
    ={}&
    \frac{(-1)^n}{4}e^{i\phi_{B_n}-z_n/2}
    \Bigg\{
    -B_n\,\mathbf n_{\mathbf x}\cdot(\mathbf x-\mathbf y)
    \left[L_{n-1}^{(1)}(z_n)+\frac12L_n(z_n)\right]
    \nonumber\\
    &\hspace{3em}
    +i\left[k_n+\mathbf n_{\mathbf x}\cdot
    \bigl(\mathbf A_n(\mathbf y)-\mathbf A_n(\mathbf x)\bigr)\right]
    L_n(z_n)
    \Bigg\},
\end{align}
where $L_{-1}^{(1)}=0$, $k_n=(2n+1)/\rho$, and
$\mathbf A_n(\mathbf x)=B_n(-x_2,x_1)/2$.

The regularization process introduces unwanted behavior, as the disappearance of the identity term at $\beta(\nu_n)=0$ in \eqref{eq:mag_cfie_fredholm_reg} permits zero-field solutions with nonzero density at $\nu_n$. For example, on the unit circle at $\nu_0=\tfrac12$, inserting the density $\sigma(s)=e^{-is}$ with parametrization $\mathbf x=r(\cos t,\sin t)$ into the SLP kernel \eqref{eq:regularized_green_mag} constructs the wavefunction analogously to \eqref{eq:wav_from_u} as 
\begin{equation}
    \psi(r,t)
    =
    \frac{e^{-B_0(r^2+1)/4}}4
    \int_0^{2\pi}
    \exp\!\left[\frac{B_0r}{2}e^{i(t-s)}\right]
    e^{-is}\,ds
    =0.
\end{equation}
Hence
\begin{equation}
    \widetilde{A}_M(\nu_0)\sigma
    =(D_n+ik_0)\psi\big|_{\partial\Omega}=0.
\end{equation}
This is an apparent solution without a physical non-zero wavefunction. 

Therefore the introduced non-physical solutions must be removed from the spectrum. For Beyn's method, we do this by fixing the interior source points $\mathbf s_j\in\Omega$ and choose the elements of the probe matrix $V$ in \eqref{eq:beyn_moments} as
\begin{equation}
    V_{ij}(k)=-(D_n+ik)\widetilde G_M(\mathbf{x_i},\mathbf s_j;k),
\end{equation}
where  $\mathbf{x_i}$ are all the boundary nodes. This choice of $V_{ij}$ becomes apparent when applying the CFIE boundary operator on it \cite{BIM_magnetic_mathematical}
\begin{equation}
[A(k)^{-1}V_j(k)]_i
\approx D_{n_i}^-G_D(\mathbf x_i,\mathbf s_j;k) ,
\end{equation}

\begin{figure}[h!]
    \centering
    \includegraphics[width=0.8\linewidth]{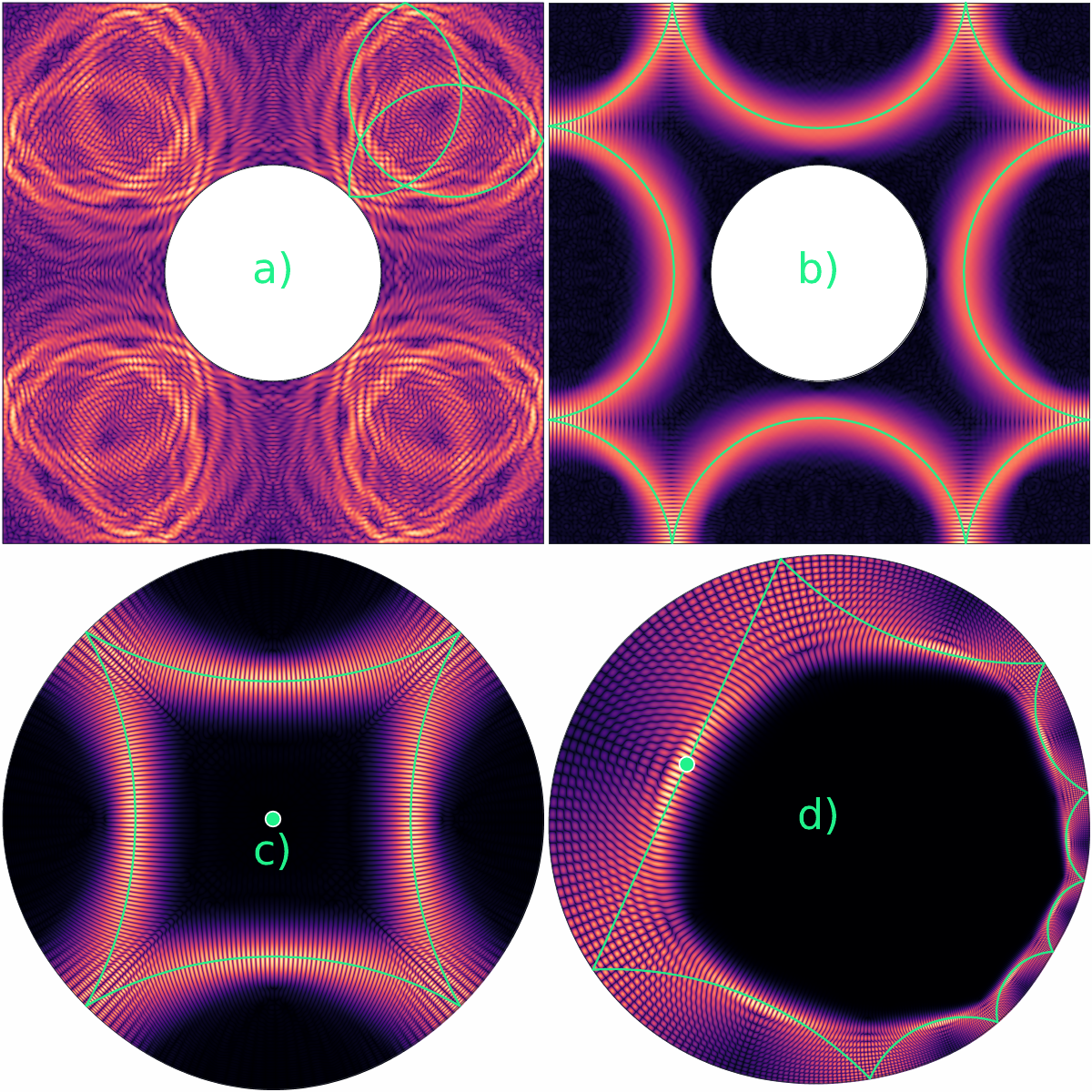}
    \caption{
    Eigenstates showing scarring near an unstable periodic orbit in panel~(a), and localization on a stable
    periodic orbit in panels~(b)--(d), in magnetic and hyperbolic billiards. The corresponding classical
    periodic orbits are overlaid in green. All physical boundaries impose the Dirichlet condition $\psi=0$.
    \textbf{(a,b)} Magnetic Sinai billiard: its geometry is defined by the \texttt{SinaiBilliard} struct where the outer boundary is $(-1,1)^2$ with a centered circular obstacle of radius $0.4$.
    \textbf{(a)} $k=218.04185826+1.66\times10^{-12}i$, $\rho=0.38$: an eigenstate scarred
    near an unstable three-bounce PO. The full eigenfunction is shown, with only the upper-right orbit overlaid.
    \textbf{(b)} $k=220.27744369+5.41\times10^{-12}i$, $\rho=0.55$: a regular state localized on a stable eight-bounce orbit involving only the outer walls.
    \textbf{(c,d)} Hyperbolic billiard with Gaussian curvature $-1$.
    Both states belong to the same physical boundary, whose centered
    Cartesian parametrization is
    $(x,y)=\tanh(r_H(\theta)/2)(\cos\theta,\sin\theta)$,
    where $r_H(\theta)=2(1+0.02\cos(4\theta))$ is the polar radius.
    \textbf{(c)} $k=99.88557156+2.19\times10^{-12}i$: regular state localized on a four-bounce PO.
    \textbf{(d)} $k=99.57879255-7.20\times10^{-13}i$:  regular state localized on an eight-bounce PO,
    with the billiard center being $(0.56977467,-0.22462278)$ (the green dot in panels \textbf{(c,d)} signifies the center of the Poincar\'e disk). One geodesic PO section passes through the Poincar\'e disk origin and therefore appears straight.
    }
    \label{fig:wav_mag_hyper}
\end{figure}

where $A$ is the Fredholm matrix of the CFIE operator, $G_D$ is the billiard Dirichlet Green function (which satisfies $G_D(x,y,k)|_{\partial \Omega}=0$), and the superscript $-$ denotes the interior normal derivative. Thus the artificial solutions induced by the $\beta(\nu)$ regularization are removable using Beyn's contour moments as by Cauchy's theorem they contribute nothing to the Beyn moments \eqref{eq:beyn_moments} since $D_{n_i}^-G_D(\mathbf x_i,\mathbf s_j;k \rightarrow k_{\nu_{n}})$ is analytic at Landau levels (if a true billiard eigenvalue is there it still gives a genuine pole there). For the intuition behind this see Appendix.~\ref{sec:half_contour}. CORK instead operates directly on the matrix polynomial $A(t)$ \eqref{eq:cork_taylor}, whose artificial Landau zeros can substantially increase the required Krylov dimension. Their removal by polynomial division in the affected columns is left to future work.

The local singularity subtraction employed in \cite{Hornberger_Smilansky} generally leaves higher-order
logarithmic terms in the numerically integrated remainder, with the choice of the  windowing function imposing additional limitations to quadrature convergence. These reduce its smoothness and can limit quadrature convergence. We instead extract the complete analytic coefficient of the logarithmic singularity, leaving a locally analytic remainder on analytic boundary segments. Here we give the simplest SLP kernel split. On a smooth boundary panel parametrized by $\mathbf x=\boldsymbol\gamma(t),\mathbf y=\boldsymbol\gamma(s)$ the regularized Green function $\widetilde G_M$ with the radial factor $g=\beta(\nu)\Gamma(\frac12-\nu)e^{-z/2}U(\frac12-\nu,1,z)$ has the logarithmic analytic split ( $\lim_{\epsilon \rightarrow0}(U(\frac{1}{2}-\nu,1+\epsilon,z)$ of Eq. 13.2.42 in \cite{NIST:DLMF})
\begin{equation}
\begin{gathered}
    g(t,s;k)=-2K_1(t,s;k)\log|t-s|+K_2(t,s;k),\\
    K_1=\beta(\nu)e^{-z/2}{}_1F_1(\tfrac12-\nu;1;z),
    \qquad
    z=\frac{k}{2\rho}|\boldsymbol\gamma(t)-\boldsymbol\gamma(s)|^2.
\end{gathered}
\label{eq:analyt_split_mag}
\end{equation}

\begin{figure}[h!]
    \centering
    \includegraphics[width=1.0\linewidth]{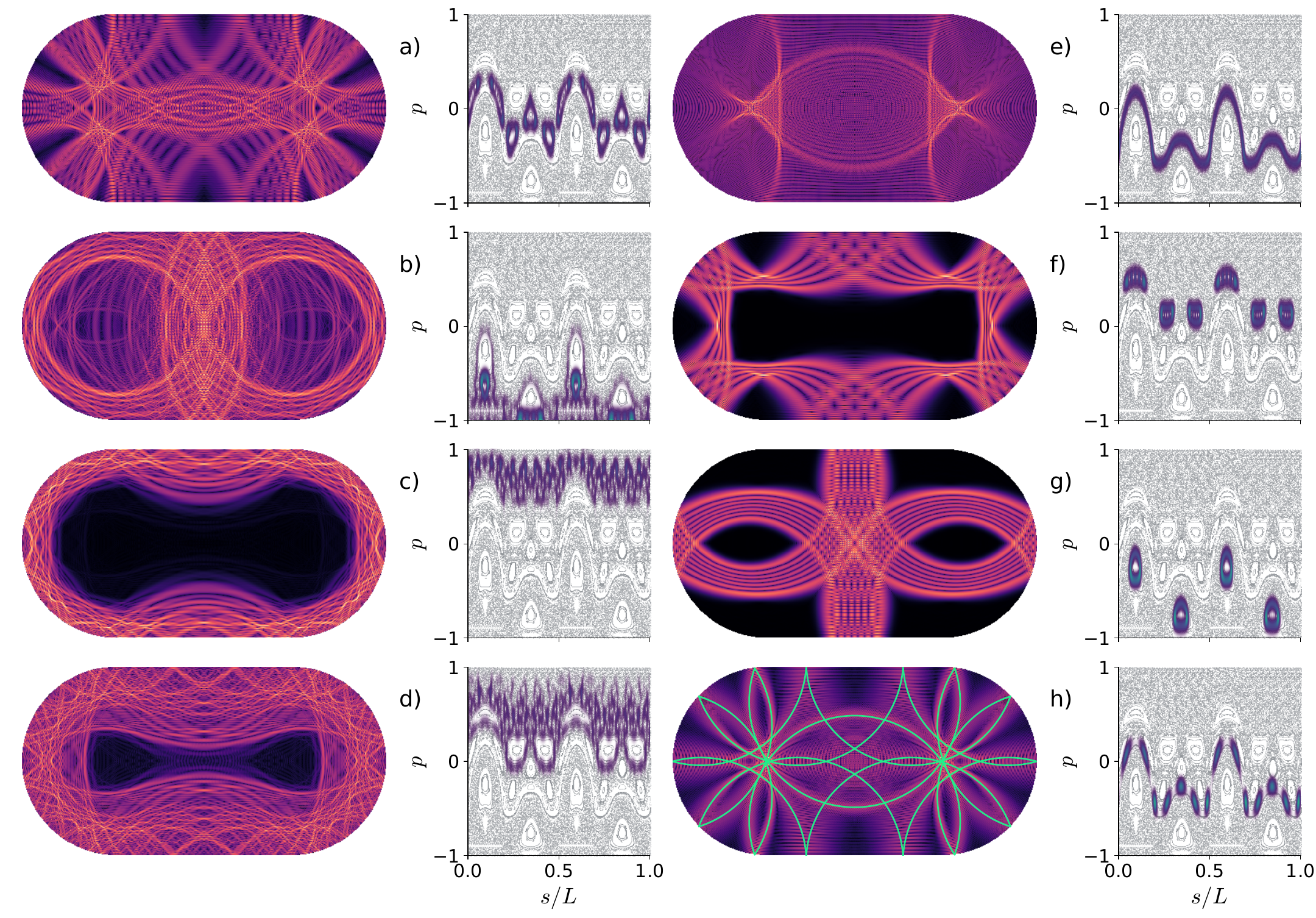}
    \caption{left: Gallery of eigenstates of the \texttt{StadiumBilliard} with half-length of the straight edge being $1.0$ and radius of the semicircles being $R=1.07$, right: their corresponding PH functions. The cyclotron radius was chosen to be $\rho=\frac{k}{B}=1.2$. All eigenstates except $b)$,$c)$ and $d)$ are regular (state $a)$ could be mixed). $b)$ is a mixed-type state, with significant overlap between the chaotic and regular regions (as seen from its PH function), while $c)$ and $d)$ are chaotic. State $h)$ is overlaid in green with a classical $14$-bounce PO.
    The eigenvalues obtained using \texttt{BeynSolver} are approximately
    \(300.84210\), \(300.85644\), \(300.85745\), \(300.87821\),
    \(300.87040\), \(300.87234\), \(300.86489\), and \(301.00944\),
    corresponding to labels (a)--(h), respectively. Their imaginary
    parts have magnitudes below \(2.1\times10^{-12}\).}
    \label{fig:stadium_mag_PO}
\end{figure}

where and $K_1,K_2$ extend smoothly across $t=s$ and ${}_1F_1$ is Kummer's confluent hypergeometric function \cite{NIST:DLMF}. For fixed $\nu(k)$ away from Landau levels, as $z\to+\infty$ (Eq. 13.2.(iv) in \cite{NIST:DLMF}),
\begin{equation}
    {}_1F_1\!\left(\frac12-\nu(k);1;z\right)
    \sim
    \frac{e^z z^{-\nu(k)-1/2}}{\Gamma(\frac12-\nu(k))}.
    \label{eq:log_coeff}
\end{equation}
Consequently, the logarithmic coefficient grows while the physical radial factor decays (Eq. 13.7.3 in \cite{NIST:DLMF} and \cite{Hornberger_Smilansky}):
\begin{align}
    K_1
    &\sim
    \frac{\beta(\nu(k))}{\Gamma(\frac12-\nu(k))}
    e^{z/2}z^{-\nu(k)-1/2},
    \\
    g
    &\sim
    \beta(\nu(k))\Gamma\!\left(\frac12-\nu(k)\right)
    e^{-z/2}z^{\nu(k)-1/2}.
\end{align}

\begin{figure}[h!]
    \centering
    \includegraphics[width=0.75\linewidth]{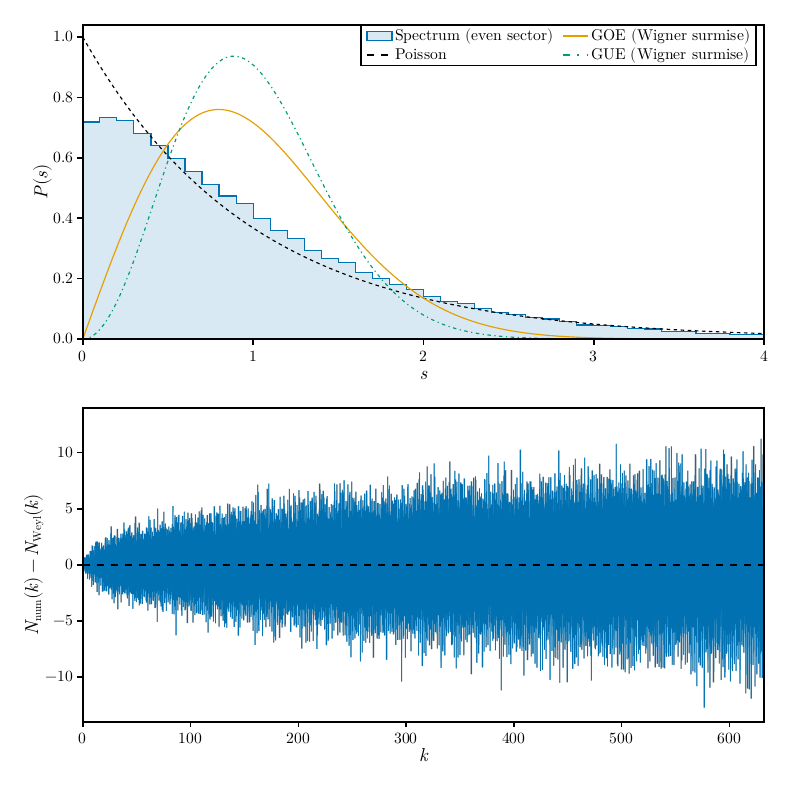}
    \caption{NNLS $P(s)$ of the unfolded levels $s$. The spectrum is of the mixed type. The unfolding is done via Weyl's law \cite{Stoe,Weyl1911,Hornberger_Smilansky}. Top row: even symmetry spectrum of the \texttt{StadiumBilliard} which has a $C_2$ symmetry when $B>0$, containing $\approx 125000$ levels. Bottom row: the oscillating part of Weyl's law of the full spectrum containing both parities ($\approx 250000$ levels), signifying no lost levels.}
    \label{fig:mag_stad_p_s}
\end{figure}

The split \eqref{eq:analyt_split_mag} therefore requires cancellation
between exponentially large terms, causing loss of accuracy and potential overflow in double precision.
The standard global periodic Kress rule applies this split over the entire closed boundary component, requiring its coefficients also at large source-target separations \cite{kress}. We instead use Gauss--Legendre logarithmic product quadrature on boundary panels \cite{SmithSloan1980_Gauss_Legendre} (efficiently using \cite{FastGaussQuadratureJL}) (convergence wrt. boundary collocation count is also being tested for the Alpert quadrature scheme \cite{alpert} since it does not require working with the analytic split \eqref{eq:analyt_split_mag} and therefore requiring no quadrature subdivision for the exponentially growing coefficient \eqref{eq:log_coeff}). On the panel containing the target, the logarithmic correction is applied only to nearby source points. Integration intervals extending beyond the turning separation $2\rho$ are subdivided, and their distant pieces are integrated using the stable full Green function \eqref{eq:regularized_green_mag}.

The accurate evaluation of $U$ is notoriously difficult for all regimes in question \cite{PearsonOlverPorter2017_Tricomi_U_difficulty}, therefore we employ stable short-step Taylor recurrences for \eqref{eq:analyt_split_mag}. Beyond the turning separation $2\rho$, the physical radial Green function decays exponentially, whereas an independent
solution grows exponentially \cite{Hornberger_Smilansky}. Both configurations will be implemented for both $\texttt{BeynSolver}$ and $\texttt{CORKSolver}$.
The eigenstates in Fig.~\ref{fig:wav_mag_hyper} were computed using CFIE \eqref{eq:cfie_operator} with Gauss-Legendre panel logarithmic product integration with Beyn's method for panel~(a) and using Kress's quadrature scheme with CORK (see Sec.\ref{sec:cork}) for panels~(b)--(d). Interior wavefunctions were evaluated using their respective single-layer potential (analogously to \eqref{eq:wav_from_u}). Fig.~\ref{fig:stadium_mag_PO} provides some more examples of billiards in a homogenous magnetic field.

Here we provide a small spectrum (Fig.\ref{fig:mag_stad_p_s}) of the \texttt{StadiumBilliard} with half-length of the straight edge being $1.0$ and radius of the semicircles being $R=1.07$ in a homogeneous magnetic field with fixed cyclotron radius $\rho=\frac{k}{B}=1.2$ (chosen such that classically no bulk states emerge \cite{Hornberger_Smilansky}), such that they underlying classical PSOS (which is mixed with these configurations) is left invariant at each energy scale. It was computed using \texttt{BeynSolver}.

\nolinenumbers

\end{document}